\documentclass[acmlarge]{acmart}

\usepackage{booktabs} 
\usepackage{algorithmic}

\usepackage[ruled]{algorithm2e} 

\usepackage{multirow}

\SetAlFnt{\small}
\SetAlCapFnt{\small}
\SetAlCapNameFnt{\small}
\SetAlCapHSkip{0pt}
\IncMargin{-\parindent}

\setcopyright{usgovmixed}

\acmDOI{0000001.0000001}

\begin{document}
\title{Fullie and Wiselie: A Dual-Stream Recurrent Convolutional Attention Model for Activity Recognition} 

\author{Kaixuan Chen}
\authornote{This is the corresponding author}
\orcid{1234-5678-9012-3456}
\affiliation{%
  \institution{University of New South Wales}
  \city{Sydney}
  \country{AU}}
\email{kaixuan.chen@student.unsw.edu.au}

\author{Lina Yao}
\affiliation{%
  \institution{University of New South Wales}
  \streetaddress{K17, UNSW}
  \city{Sydney}
  \state{NSW}
  \country{Australia}}
\email{lina.yao@unsw.edu.au}

  \author{Tao Gu}
\affiliation{%
  \institution{RMIT University}
  \city{Melbourne}
  \state{VIC}
  \country{Australia}}
 
    \author{Zhiwen Yu}
\affiliation{%
  \institution{Northwestern Polytechnical University}
  \city{Xi'an}
  \state{Shaanxi}
  \country{China}}

    \author{Xianzhi Wang}
\affiliation{%
  \institution{University of New South Wales}
  \city{Sydney}
  \state{NSW}
  \country{Australia}}
  
      \author{Dalin Zhang}
\affiliation{%
  \institution{University of New South Wales}
  \city{Sydney}
  \state{NSW}
  \country{Australia}}

\begin{abstract}
Multimodal features play a key role in wearable sensor based Human Activity Recognition (HAR).
Selecting the most salient features adaptively is a promising way to maximize the effectiveness of multimodal sensor data.
In this regard, we propose a "collect fully and select wisely (Fullie and Wiselie)" principle as well as a dual-stream recurrent convolutional attention model, Recurrent Attention and Activity Frame (RAAF), to improve the recognition performance.
We first collect modality features and the relations between each pair of features to generate activity frames, and then introduce an attention mechanism to select the most prominent regions from activity frames precisely. The selected frames not only maximize the utilization of valid features but also reduce the number of features to be computed effectively. We further analyze the hyper-parameters, accuracy, interpretability, and annotation dependency of the proposed model based on extensive experiments. The results show that RAAF achieves competitive performance on two benchmarked datasets and works well in real life scenarios.
\end{abstract}

%
%

%
%



\maketitle


\section{Introduction}

Human Activity Recognition (HAR) plays a key role in several research fields. It has gained broad attention due to the increasing popularity of ubiquitous environments, especially in health care and surveillance domains \cite{anguita2013public, zhang2012usc}.
Generally, HAR diverges into two categories of approaches: vision-based activity recognition \cite{wang2017modeling} and sensor-based activity recognition \cite{chen2012sensor}.
The sensor-based approach has several advantages over the vision-based approach and has seen diverse applications including health monitoring and motion sensing games.

\begin{itemize}
\item Compared with cameras, wearable sensors are not usually confined by environment constraints such as illumination, point of views, and set up cost. \cite{bulling2014tutorial}.
\item Sensor data obtained from wearable devices typically appear higher quality, and complicated feature extraction is not necessary as compared to image data.
\item Wearable sensors only detect the data that are strongly related to the dynamics of human motions. Therefore, sensor data collected do not violate human privacy while image data do.
\end{itemize}

Despite a large number of sensor-based recognition solutions being proposed over the decade, we discover several limitations.
First, there is still a lack of comprehensive model representation to sensor signals in a way that different activities can be distinguished in a more expressive and effective ways.
With the recent advances in deep neural networks and the notable performance achieved by these methods in the community of HAR \cite{guo2016wearable, ha2015multi},
Convolutional Neural Network (CNN) appears to be a promising candidate for building such models.
However, while CNN does well in capturing spatial relationships of features, it focuses merely on the features covered by the convolutional kernels but overlooks the correlation among non-adjacent features \cite{kavukcuoglu2010learning}.
Considering that most of the data collected by wearable sensors such as accelerometers and gyroscopes are tri-axis, in this paper, we transform sensor signals into a new activity frame which not only captures the relationships between each pair of tri-axis signals but also contains the relationship between each pair of single signals. The experiments show that our new representation is far more discriminative than traditional representations.

Second, the demerits of interperson variability and interclass similarity can greatly reduce system performance. \cite{bulling2014tutorial}.
Interperson variability comes from the fact that the same activity can be performed differently by different people, and interclass similarity results from the similarity in the behavior patterns of different activities like walking and running. Both the above issues require the classifier to be task dependent, i.e., it should automatically extract the salient information indicative of the true activity and ignore the interclass similarity. To this end, we propose an attention based model, which is directly related to the HAR task, to address the problems of interperson variability and interclass similarity. 

Attention is originally a concept in biology and psychology that implies focusing the power of noticing or thinking on something special to achieve better cognitive processes.
The attention mechanisms have several advantages, the first being task dependence. Intuitively, the motion of different body parts has varied contributions to different activities \cite{wang2017modeling, yacoob1998parameterized}. For example, jumping mostly involves legs while running is related to both arms and legs. More specifically, recognizing the patterns of walking depends more on the acceleration of legs while distinguishing sitting from lying would rely more on the orientation. 
In this paper, we separate the data related to each body part to different modals, namely accelerometer data, gyroscope data and magnetometer data, respectively. With the help of activity frames, we can analyze not only the independent modals but also their correlations thoroughly. Here, the attention mechanisms ensure that the system only focuses on the most contributing data and ignores the irrelevant sensors or modals.

The second advantage of the attention mechanisms is that it opens the black box of deep neural networks to a certain degree.
While the inner mechanisms of  neural networks remain implicit, interpretable neural network is becoming another trend in the machine learning and data mining fields. Taking convolutional neural networks for example, when using convolutional neural networks to recognize a dog from an image, we tend to explicitly know that one filter distinguishes the dog head and another filter identifies the dog paw.
Back to activity recognition, the attention model not only provides the specific body parts it focuses on but also highlights the most contributing sensors and modals to distinguish diverse activities. The salient sensor data can be inferred from the glimpse patch (to be detailed in Section~\ref{sec:Glimpse Network} ).

The third advantage is that it reduces the computational cost significantly.
Usually, the dimension of the features expands as we extract the full spatial relationships among sensors, and the cost increases with the increase of input data dimension.
Most existing models process the entire data every time, resulting in high computational cost.
Some works \cite{ooi2016image, lai2014multilinear, wold1987principal} aim to limit the input dimension using techniques such as dimensionality reduction and feature selection. However, feature processing comes with information loss, leading to a new trade-off  problem between accuracy and cost.
Inspired by human attention, our proposed method focuses on only one small patch of the data each time and goes to the next patch when necessary.
This method considerably reduces computational cost as well as information loss. 

In this paper, we tackle the HAR problems by transforming wearable sensor data into activity frames and deploying a dual-stream recurrent convolutional attention model, including one attention stream and one activity frame stream, to recognize activities. The main contributions of this work are summarized as follows:

\begin{itemize}

\item We transform the tri-axis sensor data into activity frames to extract the full relationships between data pairs. This enables the convolutional neural network to cover all features without overlooking any relationships between data pairs. Furthermore, the activity frames are encoded into convolutional activity frames in order to extract high-level features. Our model uses a single convolutional layer to encode low level data. This layer is simple yet generates an effective representation to characterize the local salience of the sensor data.
\item We propose a dual-stream recurrent model including one attention stream and one activity frame stream to recognize activities. Firstly, the system focuses on only a small patch of the activity frame that contains the most salient information to avoid unnecessary cost on less important areas, by leveraging the recurrent attention model and combining reinforcement learning. Secondly, we deploy a long short-term memory network to exploit spatial and temporal information in time-series signals and capture the dynamics of the sensor data.
\item We examine our model on two public benchmarked datasets PAMAP2 \cite{reiss2012introducing,reiss2012creating} and MHEALTH \cite{banos2014mhealthdroid,banos2015design} and perform extensive comparison with other methods, as well re-examine our approach on a new dataset collected in the real world named MARS. The experimental results show that our proposed model consistently outperforms a series of baselines and state-of-the-arts over three datasets. 

\end{itemize}

The remainder of this paper is organized as follows. Section \uppercase\expandafter{\romannumeral2} introduces the existing wearable sensor based HAR methods and attention based models briefly . Section \uppercase\expandafter{\romannumeral3} details the proposed model. Section \uppercase\expandafter{\romannumeral4} evaluates the proposed approach and compares it with state-of-the-art methods on two public datasets and one new dataset collected in the real world. In this section, we will analyze the experimental results in light of the accuracy, interpretability, latency and annotation dependency as well. Section \uppercase\expandafter{\romannumeral5} summarizes this paper.

\section{Related Work}

In this section, owing to the prevalence and outstanding performance of deep learning for HAR in recent years, we aim at giving a comprehensive review of the existing work related to deep learning for human activity recognition. Also, we briefly introduce attention mechanisms used in previous works to study salient features.

\subsection{Deep Learning for Human Activity Recognition}

Wearable sensor based human activity recognition is essentially a problem of projecting low-level sensor data to high-level activity knowledge. In our work, one basic challenge behind the "collect and select" principal is how to deeply extract features adaptive to the classification tasks and obtain the most discriminative representations. Some works employ traditional machine learning methods working on heuristic hand-crafted features \cite{bengio2013deep, yang2015deep}, which not only requires domain knowledge about activity recognition but also may potentially lead to critical limitations like error-prone bias that hinders the performance. Recently, since deep learning has embraced massive success in many fields \cite{lecun2015deep}, a flurry of research has emerged providing deep learning based solutions to various heterogeneous human activity recognition problems. The state-of-the-art deep learning based methods have made tremendous progress in improving recognition performance and widely used in either feature extraction or classification process of HAR. The rationale of the evolution is that deep learning is able to automatically extract adaptive features and spare the effort on manually extracting features and designing classifiers in details. 

Enlightened by the work done in \cite{deng2014tutorial}, we group the deep learning algorithms for human activity recognition into two categories: generative deep architectures including deep belief network, restricted Boltzmann machine and autoencoder, and discriminative deep architectures containing convolutional neural network and recurrent neural network. We will overview the recent representative works as follows. 

\subsubsection{Generative Deep Architectures}

Some existing deep learning based activity recognition solutions utilize generative deep architectures for feature extraction and deriving more discriminative representations. One of the most widely used architectures is autoencoder. To briefly demonstrate, autoencoder is usually a simple 3-layer neural network where the output units are directly related to input units and back feeds a latent representation of the input. The motivation of autoencoder is to study higher-level representation that omits noise and enhance effective information. In \cite{li2014unsupervised}, Li et al. propose to learn features by using sparse autoencoder that adds sparse constraints, that is, KL divergence to achieve better performance in activity recognition. Wang et. al \cite{wang2016human} adopt greedy pretraining to stacked auto encoder and  integrate the feature extraction process and the classifier into an architecture to jointly train them by fine-tuning parameters.

Another widely used generative deep architecture is Restricted Boltzmann Machine (RBM) \cite{hinton2006fast}. RBM shares a similar architecture with autoencoder. The difference lies in that it uses a stochastic approach. To illustrate, it uses stochastic units with specific distributions such as Gaussian or binary distribution instead of deterministic activation functions. The authors in \cite{plotz2011feature} firstly propose to deploy RBM to study feature representations for activity recognition. Inspired by this, a sequence of works take RBM as a measure to extract features for HAR. For example, \cite{fang2014recognizing} tend to exploit improving training process for RBM. They utilize contrastive gradient to fine-tune the parameters and accelerate training. \cite{lane2015deepear} employs Gaussian layer for the first layer of their RBM model and binary for the rest. Furthermore, \cite{radu2016towards} considers multimodal sensor data and designed a multimodal RBM so that each modality has an individual RBM.

Generative deep architecture enjoys the merits of unsupervised learning and high-quality representations. However, it leads to unwanted pretraining while our target is to construct an end-to-end model. Compared with this, discriminative deep architectures are more applicable and popular in previous works.

\subsubsection{Discriminative Deep Architectures} 
Discriminative deep architectures distinguish patterns by calculating the posterior distributions of
classes based on annotated data \cite{deng2014tutorial}. Existing
research can be categorized into two main directions: convolutional neural network and recurrent neural network.

According to \cite{lecun2015deep}, the theories behind convolutional neural network including sparse interactions, parameter sharing and equivariant representations .
Usually, the convolutional neural network contains (a) convolutional layers that create convolution kernels which is convolved with the layer input over a single spatial dimension to produce a tensor of outputs; (b) rectified linear unit (ReLU) layers that apply the non-saturating activation function to increase the nonlinear properties of the decision function and of the overall network without affecting the receptive fields of the convolution layers and (c) max pooling layers that down-sample the input representation, reducing its dimensionality and allowing for assumptions to be made about features contained in the sub-regions binned.
After these, there are usually (d) fully-connected layers
which perform classification or regression tasks and CNNs can learn hierarchical representations or high-performance classifiers. 

For HAR, stemming from the time-series characteristics, CNN can be used with 1D convolution and 2D convolution to combine temporal information.  1D convolution treats each axis of sensor data as a channel, flattens and unifies the outputs of each channel to be one. One example is \cite{zeng2014convolutional}, the authors proposed to treat each axis of the accelerometer as one channel and conduct the convolutional process individually. 
On the contrary, 2D convolution transforms the input into 2D matrices and considers them as images. In \cite{ha2015multi}, Ha et al. simply generate data images by combining all axis data. After that, the authors in \cite{jiang2015human} additionally consider temporal information and yield 2D time series images. 
Furthermore, \cite{singh2017transforming} harnesses
multimodal sensor data that integrates pressure sensor data and performs 2D convolutional neural network.

However, these works require massive domain knowledge when conducting transformation, which is not feasible in more general situations, compared with which, the activity frames proposed in this work not only considers temporal information and fully extracts spatial relations but also is applicable to most of multimodal sensor data with better generalization and adaptivity.

Recurrent neural network (RNN) has been proved to be effective in the fields that contains significant temporal information such as speech recognition and natural language processing, which is also the reason why RNN is applicable to HAR. Different from CNN which only takes single vector or matrix as input, RNN requires to input a sequence of vectors or matrices while each sequence has one corresponding class label. With each recurrent layer considering both the output of the previous layer and the input vector or matrix at the current layer, RNN thoroughly analyzes the sequences step by step.
To achieve better performance, LSTM (long-short
term memory) cells are introduced and usually combined with RNN.
Some previous works utilize RNN for in HAR fields \cite{Guan:2017:EDL:3120957.3090076}. In spite of the competitive performance, the time consumption and computational cost
have caused concern. To adapt RNN to HAR field where instantaneity is an important issue for developing real application, \cite{inoue2016deep} proposed a new model to which can perform RNN for HAR with high efficiency.
\cite{edel2016binarized} proposed a binarized-BLSTM
RNN model to simplify all the parameters, input, and output
to be binary to save the consumption.

In this paper, we innovatively propose a dual-stream recurrent neural network which not only considers temporal information as conventional works but also leverages attention mechanisms which are introduced next.

\subsection{Attention Mechanisms}

In our work, except for conventional deep learning approaches including convolutional neural networks and recurrent neural networks, we also resort to attention mechanisms to facilitate to select the most salient features.

Tracing back the history of selecting effective regions using attention mechanisms or similar theories, some works in the field of computer vision \cite{alexe2012searching, denil2012learning, larochelle2010learning} formulate the process of selecting as a sequential decision task. In these works, the systems decide where to focus on step by step based on the previous decisions and the whole environment. \cite{butko2008pomdp} constructs a policy gradient formulation to simulate eye movement. The authors formulate eye-move control as a problem in stochastic optimal control based on a model of visual perception. However, the too strict constraints on RNN limit the performance. \cite{denil2012learning, larochelle2010learning} further combine attention mechanisms with deep learning algorithms. \cite{denil2012learning} selects forveated images by controlling the location, orientation, scale and speed of the attended object. To minimize the selecting uncertainty, they proposed a decision-theoretic probabilistic graphical model based on RBM.
Taking policy gradient formulations and deep learning into consideration , \cite{mnih2014recurrent} proposed the recurrent attention model (RAM) for image classification with a formulation similar to \cite{butko2008pomdp} but less restrictive and leverages RNN as well. Inspired by \cite{mnih2014recurrent}, we propose a dual-stream recurrent convolutional attention model. So far, to the best of our knowledge, our work is the first one to introduce attention mechanisms to the HAR field. As feature relations are fully extracted and represented in activity frames, attention based model wisely selects salient regions to perform activity recognition.

\section{Our Model}

To fully collect effective information and wisely select salient features, our model contains two parts: (a) feature extraction
 to firstly transform wearable sensor data into 2-D matrices and use the convolutional layer to derive higher-level features. (b) a dual-stream recurrent model including one attention stream and one activity frame stream for activity recognition. Our attention stream recurrent model simulates the procedures of human brains processing visual information within several glimpses. In addition, we introduce reinforcement learning to decide which part of the activity frames it should glimpse next. The other stream is activity frame stream. Owing to the facts that the activity recognition largely depends on temporal information and that activity frames naturally capture serial relations. Activity frame based model and are more suitable for our scenarios.

\begin{figure}[htbp] 
\centering
\includegraphics[width=6.0in]{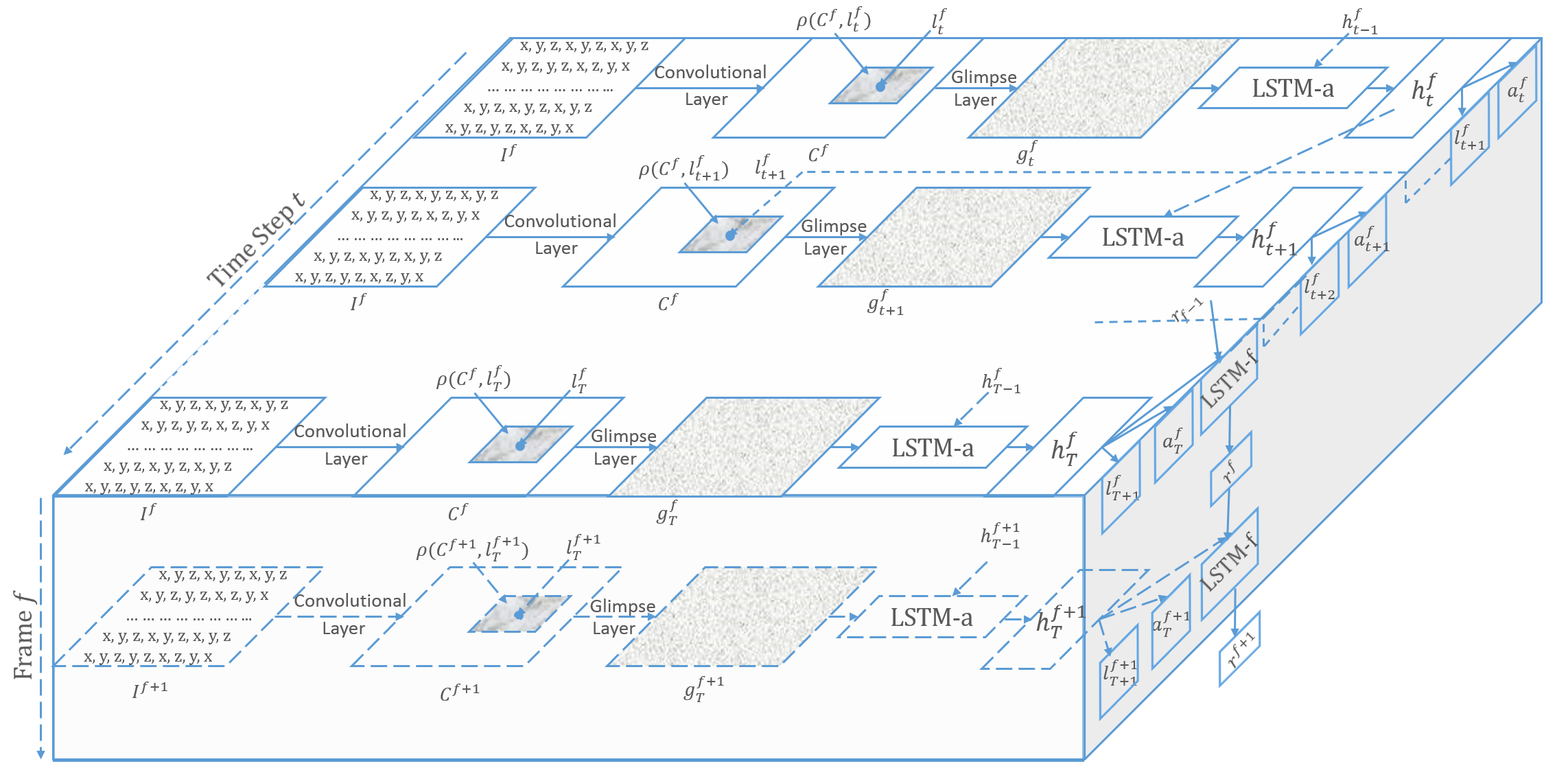} 
\caption{Work-flow of the Proposed Approach. Dashed arrows indicate the time step $t$ for attention stream and the frame $f$ for activity frame stream, respectively. For each time step $t$, the input frame goes through a convolutional layer to obtain a higher-level representation $C^f$. We extract a retina region $\rho (C^f, l^f_t)$ at location $l^f_t$, which is decided by the last time step $t-1$. $\rho (C^f, l^f_t)$  next goes through a glimpse layer to get the glimpse $g^f_t$ as input of the attention stream recurrent network, LSTM-$a$ which decides the action $a^f_t$ and the next location $l^f_t+1$. For the activity frame stream recurrent network, the LSTM-$f$ takes the last action of each frame $a_f^T$ as input and outputs the final prediction.}
\label{fig:model} 
\end{figure}

The above process is presented as a three-dimensional model in Figure~\ref{fig:model}, where the time step $t$ and frame $f$ represent the attention stream and the activity frame stream in our dual-stream method, respectively.

\begin{figure}[htbp] 
\centering
\includegraphics[width=6.0in]{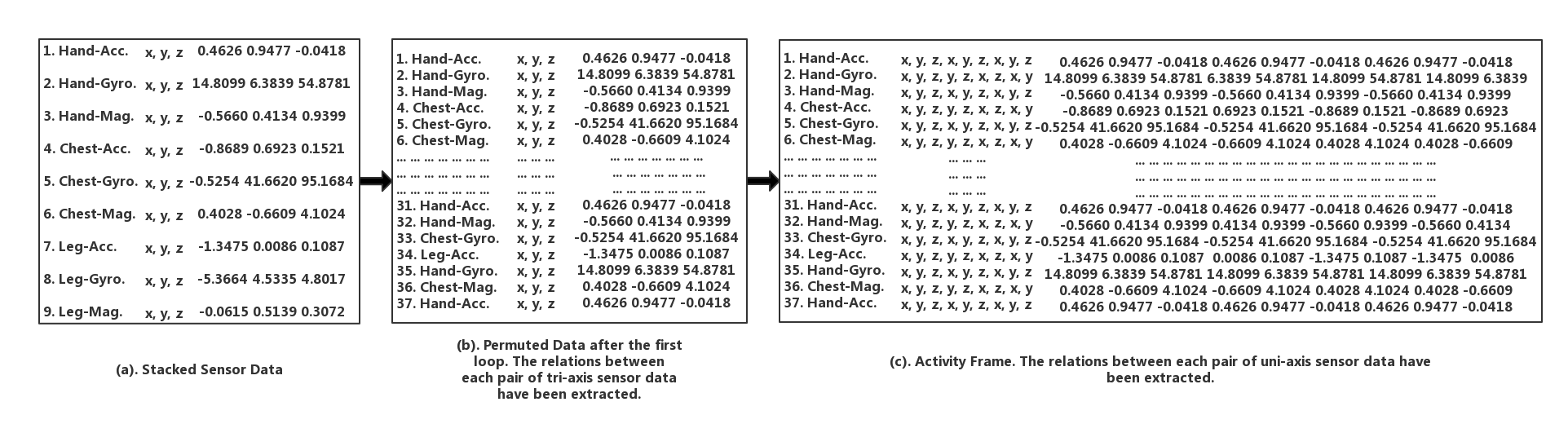}
  \caption{Transformation from sequences to frames}
  \label{fig:activity image}
  \vspace{-0.5cm}
\end{figure}

\begin{figure}[htbp] 
\centering
\includegraphics[width=6.0in]{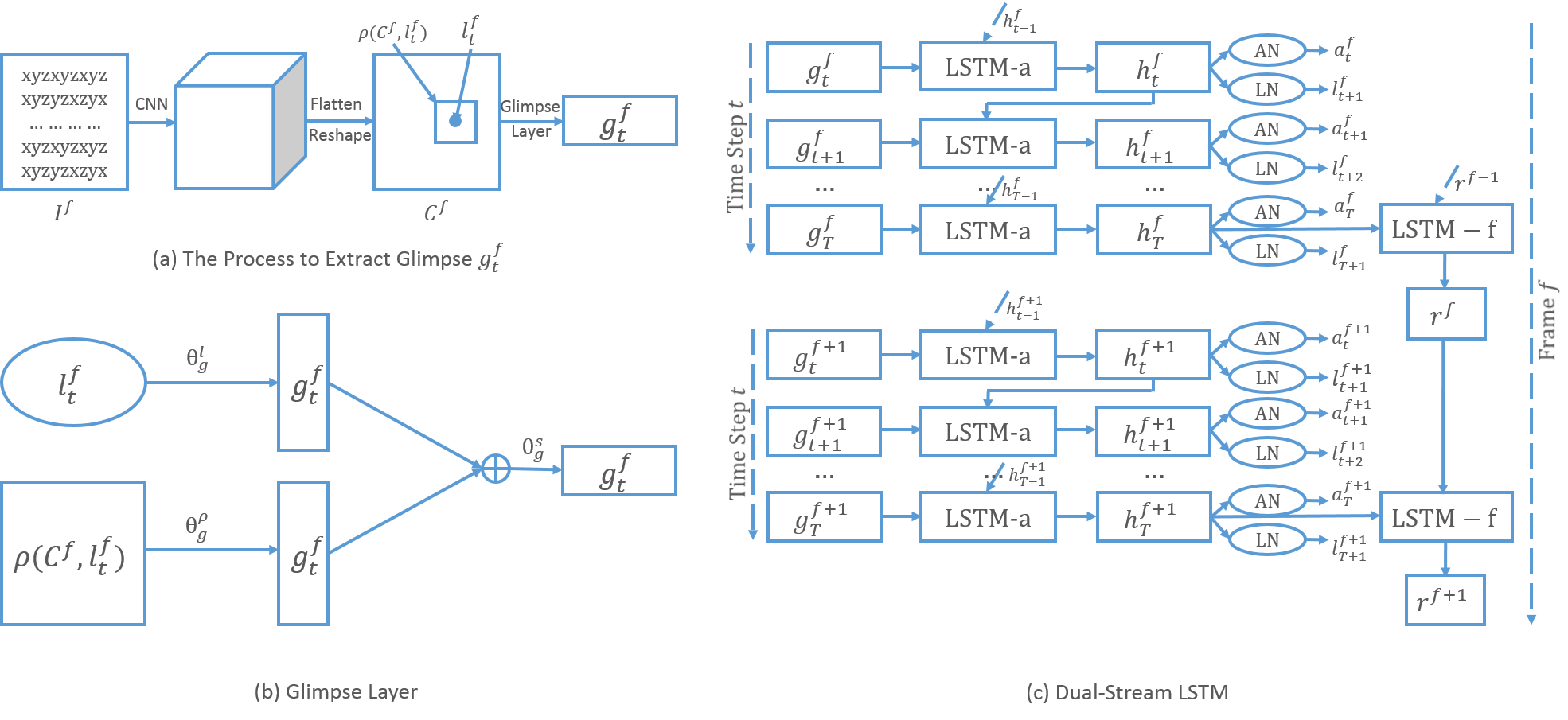} 
\caption{Flattened Model. (a) Extracting glimpse $g^f_t$ from the input activity frame, including a CNN, flattening and reshaping, and a glimpse layer. (b) The detailed description of the glimpse layer which combines the location $l^f_t$ and the retina region $\rho (C^f, l^f_t)$. (c) Dual-stream recurrent procedure containing attention stream LSTM-$a$ and activity frame stream LSTM-$f$.}
\label{fig:flattened model} 
\end{figure} 
\subsection{Input Representation}

As we transform the wearable sensor data into activity frames, the data are represented as three-dimensional vectors. Each sample $(\textbf{x, y})$ of the model consists of a 3-d vector $\textbf{x}$ and the activity label $\textbf{y}$.
Suppose  $X, Y, F$ denote activity frames' width, height, and number of frames, and $C$ represents the number of activity classes, we have:
\begin{equation}
\label{eqn:x}
\textbf{x}\in R^{X\times Y\times F}
\end{equation}

and

\begin{equation}
\label{eqn:y}
\textbf{y}\in [1, ..., C]
\end{equation}

\subsubsection{\textbf{Activity Frame}} 


There already exist some previous works that combine multimodal wearable sensor data for HAR in feature level \cite{anguita2013public, yang2015deep}. For example, Kunze et al. \cite{kunze2008dealing} concatenate acceleration and angular velocity into one vector and \cite{lara2012centinela,parkka2006activity,tapia2007real} combine acceleration and other modalities including microphone and GPS data. However, these works overlook the relations among sensors which are important to activity recognition. A popular method for extracting spatial relations is deep learning methods like CNN. Although CNN is proven to perform well in HAR \cite{jiang2015human, yang2015deep}, the accuracy is still not that satisfactory. In fact, CNN is originally proposed for images where each pixel is only related to its adjacent pixels and this small area can be easily covered by a kernel patch of a convolutional layer. However, it is still challenging to transform features to extract relations between each signal and the related signals for HAR. In many cases of HAR \cite{wang2017modeling}, the sensor data are arranged according to the physical connection of human body parts. For example, the sensor data of hands should be adjacent to the data of shoulders and the data of shoulders should be adjacent to the data of the waist, which should be followed by the data of hips, legs, and feet. Nevertheless, in the real world, activities always depend on more than one body part. For instance, running relies on the cooperation of arms and legs. In addition, the common Inertial Measurement Unit in wearable devices usually includes a tri-axis accelerometer, a tri-axis gyroscope, and a tri-axis magnetometer, and the degree to which these sensors contribute to different activities are various. This makes it even more important to find a representative transformation to extract the relationships between each pair of tri-axis sensor signals (e.g. accelerometer data and gyroscope data) and each pair of single signals (e.g. the first dimension of accelerometer data and the second dimension of gyroscope data). 

Figure~\ref{fig:activity image} shows the transformation process into activity frames. Each figure is comprised of four parts: sequence number, sensor location (hand, chest, leg) and modality (acceleration, angular velocity...), notations (x, y, z), and real data examples.
Algorithm~\ref{alg:activity frames} further illustrates the transmigration from sequences to images. First, raw signals are stacked row-by-row as shown in Figure~\ref{fig:activity image} (a). After being permuted in the first loop (line 6-18 in Algorithm~\ref{alg:activity frames}), each tri-axis sensor data has a chance to be adjacent to each of the other sensor data as shown in Figure~\ref{fig:activity image} (b). For example, supposing $N_r = 9$, then the final $S_p$ is $[1,2,3,4,5,6,7,8,9,1,3,5,7,9,2,4,6,8,1,4,7,1,5,8,2,5,9,3,6,9,4,8,3,7,2,6,1]$. Since we still need to extract the relationships between each pair of single sensor signals, the second loop (line 19-25 in Algorithm~\ref{alg:activity frames}) ensures that each single signal has a chance to be adjacent to each of the other signals as Figure~\ref{fig:activity image} (c) shows. So far we have extracted the relationships between each pair of single sensor signals.


\begin{algorithm}[!t]
\caption{Transformation from Sequences to Images}
\label{alg:activity frames}
\begin{algorithmic}[1]
\SetAlgoNoLine
\renewcommand{\algorithmicrequire}{\textbf{Input:}}
\renewcommand{\algorithmic}{\textbf{Hyper-parameters:}}
 \renewcommand{\algorithmicensure}{\textbf{Output:}}
 \REQUIRE Stacked raw signals. Each row is a tri-axis data of a accelerometer, gyroscope or a magnetometer which can be denoted as $x, y, z$.
As shown in Figure~\ref{fig:activity image} (a), each row has a sequence number. Here the number of rows $N_r$ = 9 as an example.
 \ENSURE  The activity frame $I_A$ which is a 2-D array
\STATE $i = 1;$

\STATE $j = i + 1;$

\STATE permutation sequence $S_p = [0];$

\STATE adjacent pair set $S_ap = \emptyset;$

\STATE activity frame $I_A$ = the first row of stacked signals

\WHILE{$i \neq j$}

\IF{$j > N_r$}
\STATE     {$j = 1$\;}
\ELSIF{$(i,j) \not\in S_ap$ and $(j,i) \not\in S_ap$}

\STATE add $(i, j)$ to $S_ap$;
 
\STATE  add $j$ to $S_p$;
 
\STATE  add the $j$-th row of input data to $I_A$;
 
\STATE  $i = j;$
 
\STATE  $j = i + 1;$

\ELSE
\STATE $j = j + 1$

\ENDIF
\ENDWHILE

\FOR{each row of $I_A$}
\IF{the sequence number of this row is odd}
\STATE this row is extended as $'x, y, z, x, y, z, x, y, z'$
\ELSE
\STATE this row is extended as $'x, y, z, y, z, x, z, x, y'$
\ENDIF

\ENDFOR
\RETURN $I_A$

\end{algorithmic}
\end{algorithm}

\subsubsection{\textbf{Convolutional Activity Frames}}

To derive an effective representation of features, we further transform activity frames into convolutional activity frames. Compared with a convolutional auto-encoder \cite{haque2016recurrent}, we prefer to train the model end-to-end and omit the pretraining process, as shown in Figure~\ref{fig:flattened model}. Each activity frame $I^f$ ($f$ denoted the $f_{th}$ frame) is transformed into a three-dimensional cube, the height of which depends on the number of channels of the convolutional network. The convolutional network has two convolutional layers that learn filters which activate when it detects some specific types of features at some spatial position in the input. The output is further processed by a ReLU layer and a max pooling layer.
The former applies the non-saturating activation function $relu(\nu) = max(\nu,0)$ to increase the nonlinear properties of both the decision function and the overall network without affecting the receptive fields of the convolution layer.
The latter partitions the input image into a set of non-overlapping rectangles and outputs the maximum for each such sub-region to omit the less important features.

To obtain new convolutional activity frames, the cubes are flattened and reshaped in the same size of original activity frames by a fully connected layer. After the convolutional layer, the input frame $I^f$ is encoded to be $C^f$.
 
\subsection{Attention and Activity Frame Based Recurrent Model}

We propose a dual-stream recurrent model that incorporates both attention and frame to analyze the convolutional activity frames. Figure~\ref{fig:flattened model} shows the structure of this model, where the activity frame stream recurrent modal leverages the temporal information of sensor data and the attention stream recurrent model solves the human activity recognition problem.

Since different human body parts contribute differently in recognizing different activities, we need to guarantee that the system only focuses on the most relevant and contributing parts and data. Our dual-stream model is inpired by Mnih et al. \cite{mnih2014recurrent}, who first adopt the recurrent attention model (RAM) for image classification.
Specifically, they address the image classification problem using the basic RAM. As the problem is relatively simple with only brush strokes in images being salient and the contrast between the strokes and the black backgrounds being clear.
In contrast, analyzing activity frames in this work can be  much more complex because activity frames lack such characteristics compared with image data. Moreover, almost all sensors can detect motions during activities and sometimes even standstill is still meaningful. Since the convolutional attention frames fully extract the relationships among all feature pairs, only a part of them is salient to each certain activity. Therefore, it is natural to introduce attention mechanisms facilitating to mine effective information and minimize the negative impacts of undesirable information. To the best of our knowledge, our method is the first one to leverage the attention model to tackle the activity recognition problems.
 
Figure~\ref{fig:flattened model} shows a flattened model, which better interprets the model. Our model is comprised of a glimpse network, a recurrent attention unit, and a recurrent activity frame unit that we will introduce in the followings.

 \subsubsection{\textbf{Glimpse Network}}
 \label{sec:Glimpse Network} 
 
The first part after the convolutional layer is a glimpse network. The glimpse network not only avoids the system processing the whole data in the entirety at a time but also maximally eliminates the information loss. In our model, each frame will be "understood" within $T$ glimpses. For the transformed frame $C^f$, at each time step $t$, we simulate the process of how the human eyes work. Our model first extracts a retina region denoted by $\rho (C^f, l^f_t)$ from the input data at the location $l^f_t$ with a retina. The retina image encodes the region around $l^f_t$ with high resolution but uses a progressively lower resolution for points further from $l^f_t$. This has been proved an effective method to remove noises and avoid information loss in \cite{zontak2013separating}. 
 
In the human visual system, the retina image is converted into electric signals that are relayed to the brain via the optic nerves. Likewise, in our model, the retina image is converted into a glimpse $g^f_t$ as Figure~\ref{fig:flattened model} shows. The retina image $\rho (C^f, l^f_t)$ and the location $l^f_t$ are linear transformed independently with two linear layers of network parameterized by $\theta _g^\rho$ and $\theta_g^l$, respectively. Next, the summation of these two parts is further transformed with another linear layer parameterized by $\theta _g^s$ and a rectified linear unit. The whole process can be summarized as the following equation:
 
 \begin{equation}
\label{eqn:glimpse}
g^f_t = f_g(\rho (C^f, l^f_t), l^f_t;{\theta _g^\rho, \theta_g^l, \theta_g^s}) = relu(Linear(Linear(\rho (C^f, l^f_t)) + Linear(l^f_t)))
\end{equation}
where $Linear(\bullet)$ denotes a linear transformation. Therefore, $g^f_t$ contains information from both "what" ($\rho (C^f, l^f_t)$) and "where" ($l^f_t$).

\subsubsection{\textbf{Recurrent Attention Unit}}
 
We use the recurrent neural networks as the core to process data step by step within several glimpses and introduce an attention mechanism to ensure the system only focuses on the most relevant sensors/modals and the most contributing data. The glimpses at time steps of the recurrent attention model help visualize the contribution of sensors deployed at different body parts, thus achieving better interpretability of our model.

As Figure~\ref{fig:flattened model} shows, the basic structure of the recurrent attention unit is an LSTM-$a$ (attention stream LSTM). At each time step $t$, the LSTM-$a$ receives the glimpse $g^f_t$ and the previous hidden state $h^f_{t-1}$ as the inputs parameterized by $\theta _h$. Meanwhile, it outputs the current hidden state $h^f_t$ according to the equation:

\begin{equation}
\label{eqn:h}
h^f_t = f_g(h^f_{t-1}, g^f_t;\theta _h)
\end{equation}

The recurrent attention model also contains two sub-networks: the location network and the action network. These two sub-networks receive the hidden state $h_t^f$ as the input to decide the next glimpse location $l^f_{t+1}$ and the current action $a^f_t$.
The current action not only determines the activity label $\hat{\textbf{y}}$ but also affects the environment in some cases while the location network outputs the location at time $t+1$ stochastically according to the location policy defined by a Gaussian distribution stochastic process, parameterized by the location network $f(h_t^f; \theta _t)$. 
As it decides the next region to "look at", the location network is the principal component of the recurrent attention unit.

\begin{equation}
\label{eqn:l}
l^f_{t+1} \sim P(\cdot \mid f_l(h_t^f; \theta _l))
\end{equation}

Similarly, the action network outputs the corresponding action at time $t$ and predicts the activity label given the hidden state $h_t^f$. The action $a_t^f$ obeys the distribution parameterized by $f(h_t^f; \theta _a)$. Owing to its prediction function, the network uses a softmax formulation:

\begin{equation}
\label{eqn:a}
a^f_t = f_a(h_t^f; \theta _a) = softmax(Linear(h_t^f))
\end{equation}

\subsubsection{\textbf{Recurrent Activity Frame Unit}}
Activity recognition heavily relies on the temporal information.
Therefore, besides the single activity frames used by the aforementioned process,
we additionally leverage activity frames via a recurrent activity frame unit.
As the hidden layer $h_t^f$ of the core LSTM-$a$ contributes to predicting the action $a_t^f$ and deciding the next glimpse location $l_{t+1}^f$.
For this reason, we believe the hidden state is discriminative enough to make the final prediction for the whole system.
In particular, we design an LSTM-$f$ (activity frame stream LSTM) to combine the hidden states of all the frames at the last time step $T$ to predict the activity label and to preserve the efficiency.
Given the hidden state of the last frame, the hidden state of each frame $r^f = f_r(h_T^f, r^{f-1}; \theta _r)$, parameterized by $\theta _r$.

\subsection{Training and Optimization}

Our proposed model depends on the parameters of every components, including the glimpse network, the recurrent attention network, the two sub-networks, and the activity frame stream recurrent network $\Theta = {\theta_g, \theta_h, \theta_a, \theta_l, \theta_r}$. Both the action network and the frame-based recurrent network are based on classification methods. Therefore, their parameters, $\theta_a$ and $\theta_r$, can be trained by optimizing the cross-entropy loss and the backpropagation. However, the location network should be able to select a sequence of salient regions from activity frames adaptively. Since this network is non-differentiable owing to its stochasticity and the problem can also be regarded as a control problem to settle the attention region at the next step, it can be trained by reinforcement methods to learn the optimal policies.

We simply introduce some definitions of reinforcement learning based on our case.
\begin{itemize}
\item Agent: the brain to make decisions, which is the location network in our case.
\item Environment: the unknown world that may affect the agent's decision or may be influenced by the agent.
\item Reward: the feedback from the environment to evaluate the action. In our case, for each frame, the model gives a prediction $\hat{\textbf{y}} = a_t$ and receives a reward $r_t$ as a feedback for the future correction of the prediction after each time step $t$. Suppose $T$ denotes the number of steps in our attention stream recurrent model. $r_t = 1$ if $\hat{\textbf{y}} = \textbf{y}$ after $T$ steps and $0$ otherwise. The target of the optimization is to maximize $R = \sum_{t=1}^{T} r_t$.
\item Policy: the projection from states to actions, denoted by $\pi(a\mid s) = P[A_t=a\mid S_t=s]$. To maximize the reward $R$, we learn an optimal policy $\pi(l_t,a_t|s_{1:t};\Theta)$ to map the attention sequence $s_{1:t}$ to a distribution over actions for the current time step, where the policy $\pi$ is decided by $\Theta$ of the recurrent attention model.

\end{itemize}

Based on the above discussion, we deploy a Partially Observable Markov Decision Process (POMDP) to solve the training and optimization problems, for which the true state of the environment is unobserved. Let $s_{1:t} = \textbf{x}_1, l_1, a_1;... \textbf{x}_t, l_t, a_t$ be the sequence of the input, location and action pairs. This sequence, called an attention sequence, shows the order of the regions our attention focuses on. 

To sum up, in our case, the location network is formulated as a random stochastic process (the Gaussian distribution) parameterized by $\Theta$. Each time after the location selection, the prediction $a$ is evaluated to back feed a reward for conducting the backpropagation training process. The process is also defined as policy gradient. Our goal is to maximize the simulated rewards using gradient.

Generally, for sample $x$ with its reward $f(x)$ and the probability $p(x)$, we have:
\begin{equation}
E_x[f(x)] = \sum_{x}p(x)f(x)
\end{equation}

so that the gradient can be calculated according to the REINFORCE rule \cite{williams1992simple}:

\begin{eqnarray}
\label{eqn:R rule}
\triangledown_\theta E_x[f(x)] &=& \triangledown_\theta\sum_{x}p(x)f(x) \nonumber \\
&=& \sum_{x}\triangledown_\theta p(x)f(x) \nonumber\\
&=& \sum_{x}p(x)\frac{\triangledown_\theta p(x)}{p(x)}f(x) \nonumber\\
&=& \sum_{x}p(x)\triangledown_\theta log p(x)f(x)      \nonumber \\
&=& E_x[f(x)\triangledown_\theta log p(x)]
\end{eqnarray}

In our case, given the reward $R$ and the attention sequence $s_{1:T}$, the reward function to be maximized is as follows:

\begin{equation}
\label{eqn:J}
J(\Theta) = \mathbb{E}_{p(s_{1:T};\Theta)}[\sum_{t=1}^{T} r_t] = \mathbb{E}_{p(s_{1:T};\Theta)}[R]
\end{equation}

By considering the training problem as a POMDP, a sample approximation to the gradient is calculated as follows:

\begin{equation}
\triangledown_\Theta J = \sum_{t=1}^{T} \mathbb{E}_{p(s_{1:T};\Theta)}[\triangledown_\Theta log \pi(\textbf{y}|s_{1:t};\Theta)R]
\end{equation}
where $i$ denotes the $i^{th}$ training sample, $\textbf{y}^{(i}$ is the correct label for the $i^{th}$ sample, and $\triangledown_\Theta log \pi(\textbf{y}^{(i)}|s_{1:t}^i;\Theta)$ is the gradient of LSTM-$a$ calculated by backpropagation.

We use Monte Carlo sampling which utilizes randomness to yield results that might be deterministic theoretically. Supposing $M$ is the number of Monte Carlo sampling copies, we duplicate the same convolutional activity frames for $M$ times and average them as the prediction results to overcome the randomness in the network, where the $M$ duplication generates $M$ subtly different results owing to the stochasticity, so we have:
\begin{equation}
\label{eqn:REINFORCE rule}
\triangledown_\Theta J = \sum_{t=1}^{T} \mathbb{E}_{p(s_{1:t};\Theta)}[\triangledown_\Theta log \pi(\textbf{y}|s_{1:t};\Theta)R] \thickapprox \frac{1}{M} \sum_{i=1}^{M} \sum_{t=1}^{T} \triangledown_\Theta log \pi(\textbf{y}^{(i)}|s_{1:t}^i;\Theta)R^{(i)}
\end{equation}

Therefore, although the best attention sequences are unknown, our proposed model can learn the optimal policy in the light of the reward.

To summarize, we propose a dual-stream recurrent convolutional attention model which includes transforming features into activity frames and a dual-stream recurrent model. Firstly, to fully extract relations between each pair of sensors and modality features, the inputs are innovatively transformed into convolutional activity frames. After that, the model effectively combines attention based recurrent spatial relations and recurrent temporal information to wisely select salient features and perform classification. To further illustrate the process detailedly, an overall algorithm is shown in Algorithm~\ref{alg:Overall Process of RAAF}. The experimental results presented next show that the proposed approach outperforms the state-of-the-art HAR methods.

\begin{algorithm}[!t]
\caption{Overall Process of RAAF}
\label{alg:Overall Process of RAAF}
\begin{algorithmic}[1]
\SetAlgoNoLine
\renewcommand{\algorithmicrequire}{\textbf{Input:}}
\renewcommand{\algorithmic}{\textbf{Hyper-parameters:}}
 \renewcommand{\algorithmicensure}{\textbf{Output:}}
 \REQUIRE Activity frames from Algorithm~\ref{alg:activity frames}, $T$: the number of time steps, $F$: the number of activity frames.
 \ENSURE  The prediction results

\STATE $r_{last} = RandomInitialize()$

\FOR{$f$ from $1$ to $F$}
\STATE $I = the f_th$ activity frame
\STATE $I_{CNN} = CNN(I)$
\STATE $C = Reshape(Flatten(I_{CNN}))$
\STATE $h_{last} = RandomInitialize()$
\STATE $l = RandomInitialize()$
\FOR{$t$ from $0$ to $T$}
\STATE $\rho = ExtractRetina(C, l)$
\STATE $glimpse = relu(Linear(Linear(\rho) + Linear(L)))$
\STATE $h = LSTM_{attention}(glimpse, h_{last})$
\STATE $a = softmax(Linear(h))$ trained by cross-entropy and gradient propagation
\STATE $l = tanh(Linear(h))$ trained by equation~\ref{eqn:REINFORCE rule}
\STATE $h_{last} = h$
\ENDFOR
\STATE $r = LSTM_{frame}(h, r_{last})$
\ENDFOR
\STATE $activity\_label = r$
\RETURN $activity\_label$

\end{algorithmic}
\end{algorithm}

\section{Experiments}
In this section, we present the validation of our proposed method via experiments on on two public datasets and another real-world dataset collected by ourselves. Firstly, we describe the used dataset and the experimental setup. Secondly, we present our investigation of hyper-parameter study on the classification performance. Thirdly, we compare the accuracy of our proposed methods with several state-of-the-art HAR methods, present the confusion matrices on the datasets, and analyze the experimental results. Lastly, we show the interpretability and low dependency of RAAF on labeled data.

\subsection{Datasets and Experimental Settings}

We evaluate the proposed method on two public benchmarked activity recognition datasets, PAMAP2 dataset and MHEALTH dataset and the real-world dataset MARS which is collected by ourselves.
These public datasets are the latest available wearable sensor-based datasets with complete annotation and have been widely used in the activity recognition research community.

\textbf{PAMAP2. } The dataset was collected in a constrained setting where 9 participants (1 female and 8 males) performed 12 daily living activities including basic actions(standing, walking) and sportive exercises(running, playing soccer). Six activities were carried out by the subjects optionally. The sensor data were collected at the frequency of 100 Hz from the hardware setup that contains 3 Colibri Inertial Measurement Units (IMUs) attached to the dominant wrist, the chest and the dominant side's ankle, respectively.
Besides, heart rate (bpm) was collected by an HR-monitor at the sampling frequency of 9 Hz.
All the above collected data include two 3-axis accelerometer data ($ms^{-2}$), 3-axis gyroscope data (rad/s), 3-axis magnetometer data ($/mu$T), 3-axis orientation data, and temperature ($^{\circ}C$).
Specially, temperature is collected from 3 IMUs, so it is also processed to be 3-axis.
Our experiments only consider the high-quality part of data, including temperature, accelerometer, gyroscope, and magnetometer data, to ensure effective validation of the experimental results.

\textbf{MHEALTH. } The Mobile Health (MHEALTH) dataset is also devised to benchmark methods of human activities recognition based on multimodal wearable sensor data. Three IMUs were respectively placed on 10 participants' chest, right wrist, and left ankle to record the accelerometer ($ms^{-2}$), gyroscope (deg/s) and the magnetometer (local) data while they were performing 12 activities. The IMU on the chest also collected 2-lead ECG data (mV) to monitor the electrical activity of the heart. All sensing modals are recorded at the frequency of 50 Hz. 

\textbf{MARS.} Our new dataset, the Multimodal Activity Recognition with Sensing (MARS). MARS dataset, was collected while 8 participants (6 males, 2 females) were doing 5 basic activities (sitting, standing, walking, ascending stairs and descending stairs). Three IMU sensors, Phidget Spatial 3/3/3 \cite{phidgets20101056} were attached to the dominant wrist, the waist, and the dominant side's ankle, respectively, to collect 3-axis accelerometer data (gravitational acceleration $g$), 3-axis gyroscope data ($^{\circ}/s$), and 3-axis magnetometer data ($nT$). Since participants went up and down through the same flight of stairs during our collecting of data, the magnetometer data contain signals of two opposite directions. To avoid the misconduct resulted from the opposite data, we excluded the magnetometer data for activity recognition. All IMUs collected the data at the frequency of 70 Hz.

Similar to \cite{guo2016wearable}, the experiments conducted on the two public datasets perform background activity recognition task \cite{reiss2012introducing}. The activities are categorized into 6 classes: lying, sitting/standing, walking, running, cycling and other activities. To tackle the task and ensure the rigorousness, all experiments are performed by Leave-One-Subject-Out (LOSO) cross-validation which can also test the person independence during the evaluation. The evaluation results are measured by accuracy (\%), one of the most commonly used performance measure standards for classification tasks.

Here, we describe the common design for all the experiments but leave hyper-parameter study to the next section.

\textbf{Convolutional Network: } The convolutional network has three sections. Each of the first two sections composes of one convolutional layer with the kernel size of $3$ x $3$, one rectified linear unit (ReLU) layer that applies the non-saturating activation function $relu(\nu) = max(\nu,0)$, and one max pooling layer with the kernel size of $1$ x $3$ and the stride of $1$ x $3$. The third section has a fully connected layer developed on the flattened results of the second layer. The size of the fully connected layer depends on the size of the input activity frame for the reason that the output should be reshaped to another 2-D matrix $C^f$ with the same size of $I^f$.

\textbf{Glimpse Network: } The glimpse network has three fully connected layers defined as $gl^f_t = Linear(l^f_t)$, $g\rho^f_t = Linear(\rho (I^f, l^f_t))$, $g^f_t = relu(Linear(gl^f_t + g\rho^f_t))$, respectively. The dimensionality of $g\rho^f_t, gl^f_t$ and $g^f_t$ are $128, 128$ and $220$ in our experiments.

\textbf{Action and Location Networks: } The action network only has one fully connected layer while the policy for the location network is defined by a dual-component Gaussian with a variance fixed to be 0.22. The location network outputs the location at time $t+1$ stochastically according to the location distribution, which is defined as $l^f_{t+1} = tanh(Linear(h_t^f))$.

\textbf{Two Recurrent Networks: } The proposed method has two recurrent networks. One is the attention based LSTM with the cell size of 100. The number of time steps is 40, which defines the number of glimpses. The other one is the frame-based recurrent network which has an LSTM in a size of 1000 and the number of time steps is set to 5, which decides the number of frames that are utilized to perform the recognition task.

\subsection{Hyper-Parameter Study}
\begin{figure}[htbp] 
\centering
\includegraphics[width=6.0in]{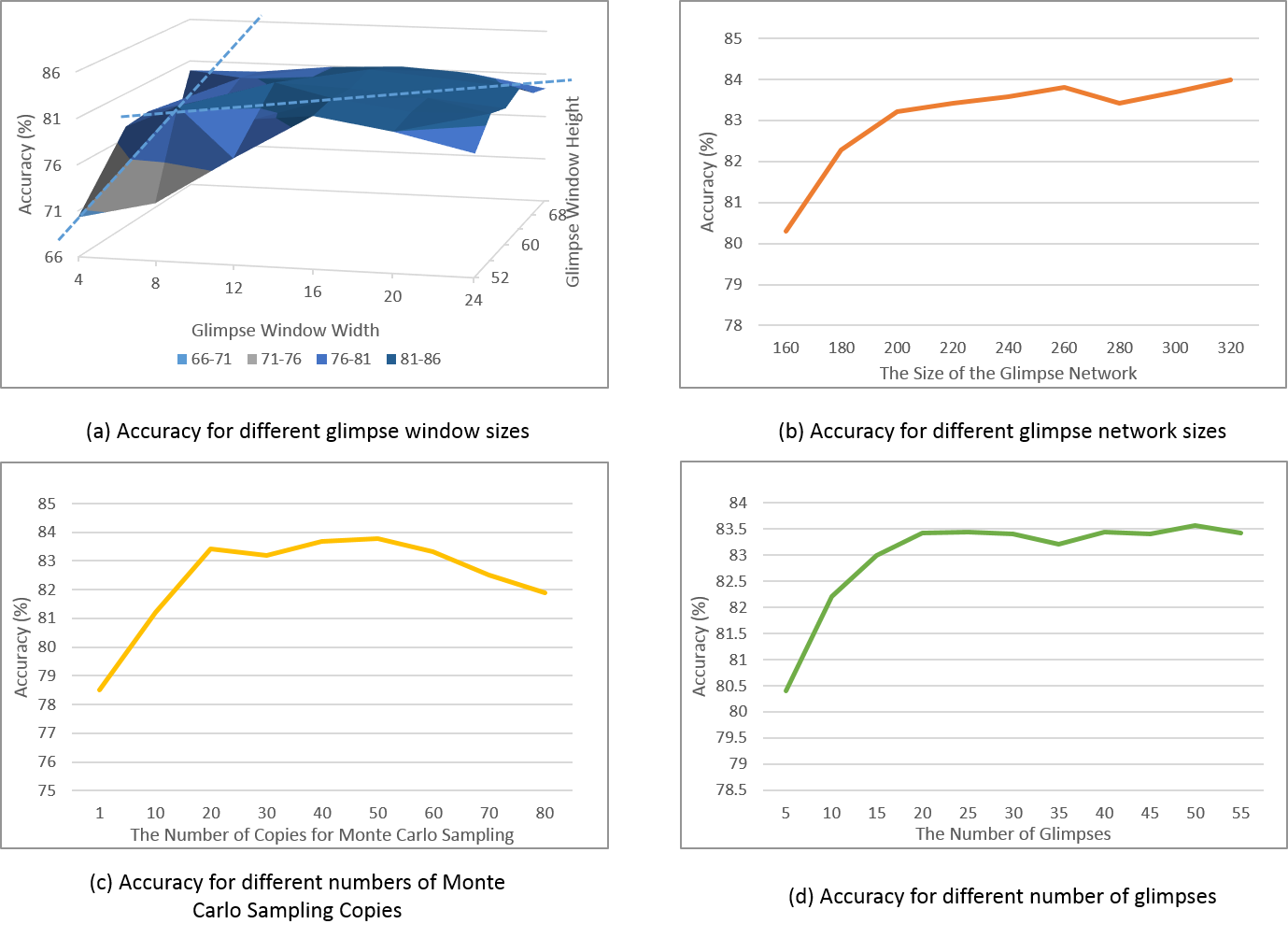} 
\caption{Experimental Results for Hyper-Parameter Tuning on PAMAP2}
\label{fig:parameter_tuning_pamap} 
\end{figure}

\begin{figure}[htbp] 
\centering
\includegraphics[width=6.0in]{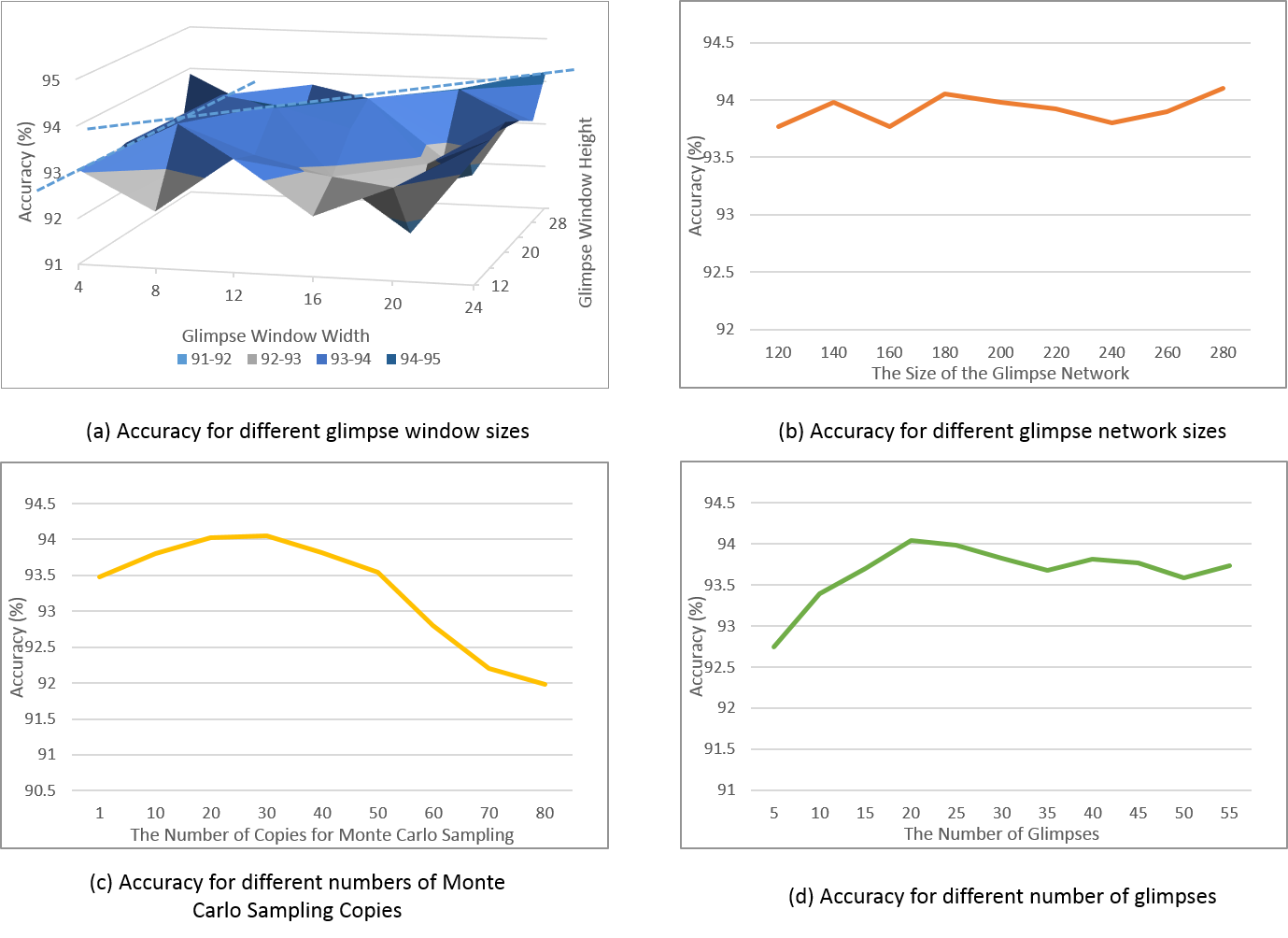} 
\caption{Experimental Results for Hyper-Parameter Tuning on MHEALTH}
\label{fig:parameter_tuning_mhealth} 
\end{figure}

\begin{figure}[htbp] 
\centering
\includegraphics[width=6.0in]{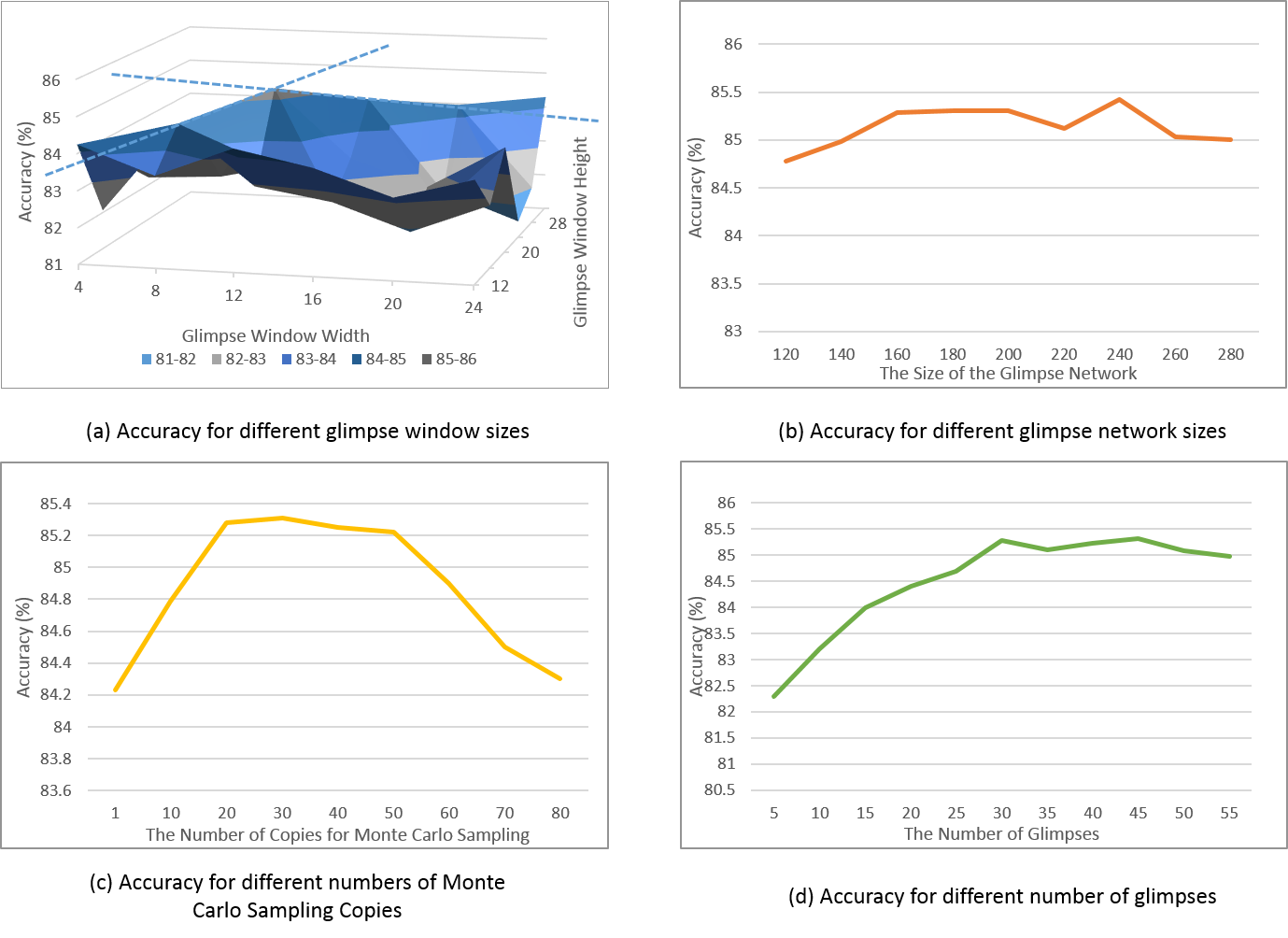} 
\caption{Experimental Results for Hyper-Parameter Tuning on MARS}
\label{fig:parameter_tuning_mars} 
\end{figure}

In this section, we mainly analyze four most contributing hyper-parameters to which the model is more sensitive in our experiments, namely the size of glimpse windows (width and height), size of the glimpse output $g^f_t$, the number of copies for Monte Carlo sampling, and the number of glimpses. For the other hyper-parameters, we just use fixed empirical values as suggested in the previous subsection. The variation trend is shown as Figure~\ref{fig:parameter_tuning_pamap}, Figure~\ref{fig:parameter_tuning_mhealth} and Figure~\ref{fig:parameter_tuning_mars}.

Taking Figure~\ref{fig:parameter_tuning_pamap} as an example, firstly, we tune the width and the height of glimpse windows to figure out their relationship as shown in Figure~\ref{fig:parameter_tuning_pamap} (a).
Specifically, there are 13 3-axis vectors to present the temperature, accelerometer, gyroscope and magnetometer data in our experiments. After Algorithm~\ref{alg:activity frames}, several $78$x$9$ activity frames are generated. Figure~\ref{fig:parameter_tuning_pamap} (a) shows that the accuracy achieves the best when the glimpse window size is $64$x$16$ and there is an obvious "ridge" along which the whole figure is almost symmetric. All the points on the symmetric line are in a ratio of $4:1$.
This suggests that the approach favors a fixed ratio of the two dimensions of the glimpse window, in spite that we used the ratio of the activity frame size of $78:9$. Also, we can see that Figure~\ref{fig:parameter_tuning_mhealth} (a) and Figure~\ref{fig:parameter_tuning_mars} (a) both show the "ridge" while their optional glimpse window sizes are different because of different sizes of activity frames.

Figure~\ref{fig:parameter_tuning_pamap} (b), (c) and (d) show the experimental results of our studies on the effect of other three hyper-parameters, the size of the glimpse network, the number of copies for Monte Carlo sampling and the number of glimpses.
In particular, Figure~\ref{fig:parameter_tuning_pamap} (b), and (d) present similar trends that the accuracy increases remarkably at first and keeps rising slowly (Figure~\ref{fig:parameter_tuning_pamap} (b)) or remains stable (Figure~\ref{fig:parameter_tuning_pamap} (d)) after getting a turning point. However, for the Monte Carlo sampling, too low or too high values lead to worse performance, as Figure~\ref{fig:parameter_tuning_pamap} (c) shows. Considering the computational complexity increases with larger values of the hyper-parameters, a trade-off between the accuracy and the computational complexity is necessary, especially for Monte Carlo Sampling. Therefore, we simply select the points slightly after the turning points (220, 20, 30) as the optimal parameters to conduct our following experiments. And we can notice that the variation trends in Figure~\ref{fig:parameter_tuning_mhealth} and Figure~\ref{fig:parameter_tuning_mars} enjoy the same patterns. 

\subsection{Accuracy Comparison and Performance Analysis}

To evaluate the performance of the proposed approach, RAAF, we conduct extensive experiments to compare its performance with the state-of-the-art methods on PAMAP2 and MHEALTH. We elaborately select other four state-of-the-art and multimodal feature-based approaches (MARCEL \cite{guo2016wearable}, FEM \cite{lara2012centinela}, CEM \cite{guo2014activity} and MKL \cite{althloothi2014human}) and five baseline methods (Support Vector Machine (SVM), Random Forest(RF), K-Nearest Neighbors(KNN), Decision Tree(DT) and Single Neural Networks) to show the competitive power of the proposed method. To ensure fair comparison, the best parameters test, RAAF, is used on both datasets; the best trade-off parameter ($\lambda = 0.7$) is deployed for MARCEL; time-domain features including mean, variance, standard deviation, median and frequency-domain features including entropy and spectral entropy are utilized for FEM; each modality feature group are defined an independent kernel for MKL; and for other baseline methods, all modality features are deployed. All parameters adopted are in reference to the parameters suggested in literature. The results in Table~\ref{tab:system_comparison} show the proposed RAAF outperforms all the state-of-the-art methods and the baseline methods.

\begin{table}[]
\centering
\caption{Comparison among RAAF and four state-of-the-art methods and five baseline methods. For PAMAP2 dataset, accelerometer, gyroscope and magnetometer data are utilized. For MHEALTH dataset, ECG data are considered additionally.}
\label{tab:system_comparison}
\begin{tabular}{cccccc}
\hline
Datasets & \multicolumn{5}{c}{Methods} \\ \hline
\multirow{4}{*}{PAMAP2} & RAAF & MARCEL \cite{guo2016wearable} & FEM+SVM \cite{lara2012centinela} & CEM \cite{guo2014activity} & FEM+MKL \cite{lara2012centinela, althloothi2014human} \\
& $\textbf{83.4}$ & 82.8 & 76.4 & 81 & 81.6\\ \cline{2-6}
& SVM & RF & KNN & DT & Single NN \\
& 59.3 & 64.7 & 70.3 & 57.8 & 72.0 \\ \hline
\multirow{4}{*}{MHEALTH} & RAAF & MARCEL \cite{guo2016wearable} & FEM+SVM \cite{lara2012centinela} & CEM \cite{guo2014activity} & FEM+MKL \cite{lara2012centinela, althloothi2014human} \\
& $\textbf{94.0}$ & 92.3 & 70.7 & 74.8 & 90.6\\ \cline{2-6}
& SVM & RF & KNN & DT & Single NN \\
& 68.7 & 82.5 & 86.1 & 78.7 & 89.1 \\ \hline

\end{tabular}
\end{table}

\begin{figure}[htp] 
\centering
\includegraphics[width=6.0in]{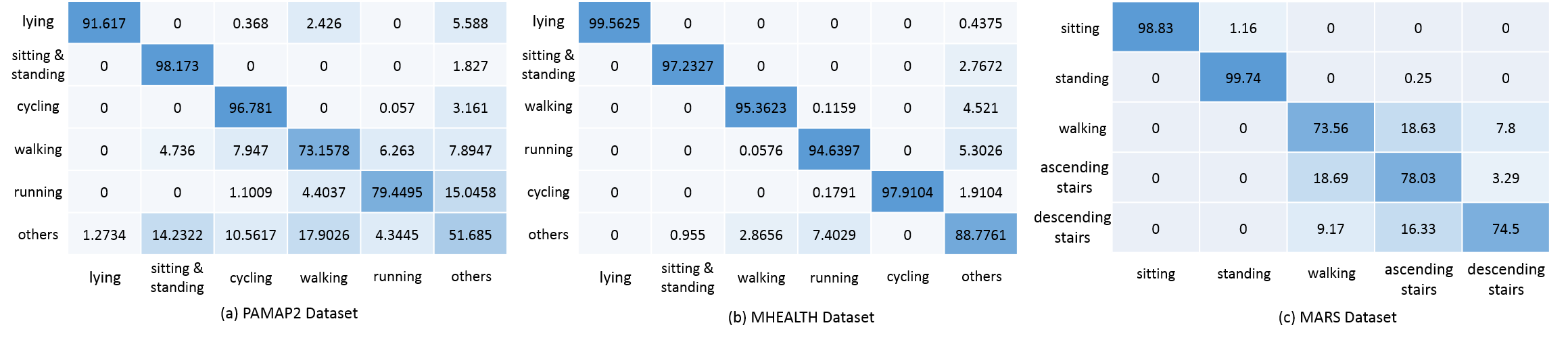} 
\caption{The confusion matrices of RAAF for background activity recognition on three datasets}
\label{fig:confusion_matrix_6_classes} 
\end{figure}

\begin{figure}[htp] 
\centering
\includegraphics[width=6.0in]{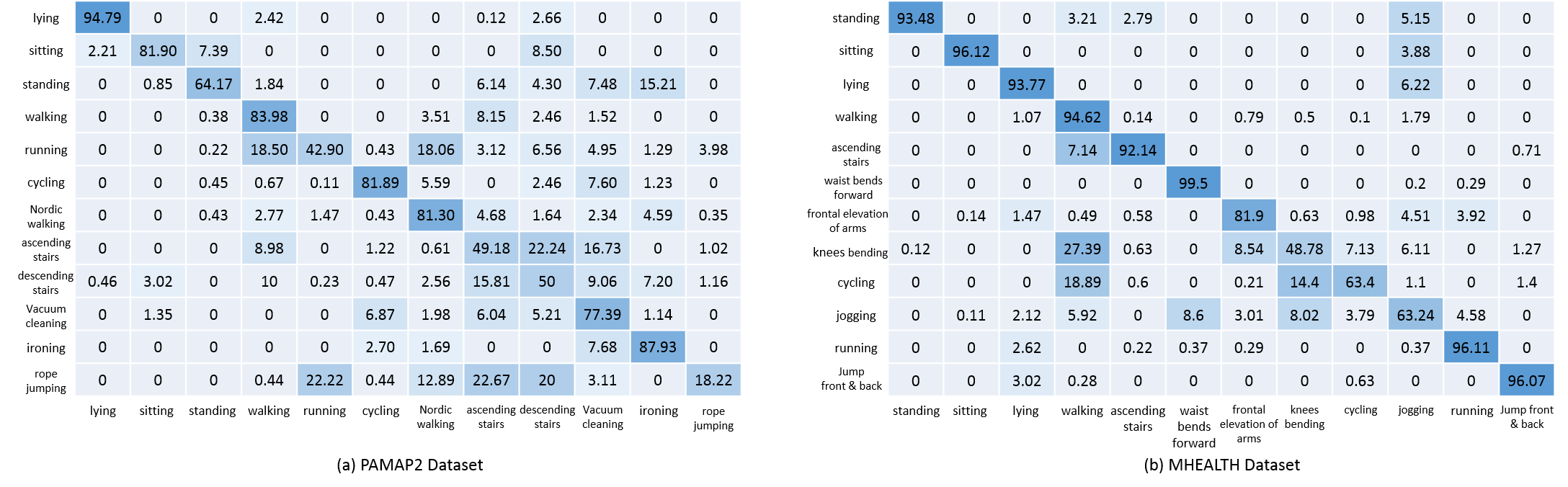} 
\caption{The confusion matrices of RAAF for all activity recognition on two public datasets}
\label{fig:confusion_matrix_12_classes} 
\end{figure}

To further explain the accuracy of RAAF on each specific activity, Figure~\ref{fig:confusion_matrix_6_classes} (a) and (b) show the confusion matrices on both public datasets performing the background activity recognition task.
The results show the proposed approach performs well for most activities such as lying, sitting and standing, and cycling.
However, more misclassifications occur on activities that have similar patterns to the background activities, such as walking, ascending stairs and descending stairs, due to the constraint of the background activity recognition task, "others". This pattern can also been seen in Figure~\ref{fig:confusion_matrix_6_classes} (c) where sitting and standing can be clearly classified while walking and ascending or descending stairs appear to be slightly confusing. To present the effectiveness of our method on other activities, Figure~\ref{fig:confusion_matrix_12_classes} shows the confusion matrices on both public datasets performing the all activity recognition task that defines separate classes for each of the 12 activities \cite{reiss2012introducing} on PAMAP2 and MHEALTH. From Figure~\ref{fig:confusion_matrix_12_classes} we observe that on PAMAP2 dataset the model works well for most activities but is confused with running, ascending \& descending stairs and rope jumping because of their similar patterns. And on MHEALTH dataset, the performance is remarkable except for some misclassifications for knees bending, cycling and jogging.

We prove the effectiveness of our activity frames by deploying the dual-stream recurrent convolutional attention model on original features. To adapt features to the proposed model, multimodal features are stacked to form original frames, as Figure~\ref{fig:activity image}(a) shows. Table~\ref{tab:feature extraction capability} presents feature extraction capability of activity frames, which shows that the proposed model based on the original frames outperforms most of the state-of-the-art methods (listed in Table~\ref{tab:system_comparison}) even without activity frames. But utilizing the activity frames can significantly improve the performance of original model due to the availability of the full relationship among features provided by activity frames.

\begin{table}[]
\centering
\caption{Feature Extraction Capability of Activity Frames}
\label{tab:feature extraction capability}
\begin{tabular}{cccc}
\hline
& PAMAP2 Dataset & MHEALTH Dataset & MARS Dataset\\ \hline
Original Frames & 81.35 & 92.20 & 77.25\\
Activity Frames & \textbf{83.42} & \textbf{94.04} & \textbf{85.28}\\ \hline

\end{tabular}
\end{table}

As latency is a critical indicator to evaluate the applicability of HAR systems in practical scenarios, table~\ref{Tab: Latency Analysis} shows the latency for testing one sample on the three datasets (all less than 1 second), which we believe is fairly acceptable in realistic application scenarios.

\begin{table}[]
\centering
\caption{Latency Analysis on Three Datasets}
\label{Tab: Latency Analysis}
\begin{tabular}{ccc}
\hline
PAMAP Dataset & MHEALTH Dataset & MARS Dataset \\ \hline
0.68s & 0.72s & 0.59s \\ \hline

\hline
\end{tabular}
\end{table}

\subsection{Model Interpretability}

\begin{figure}[htbp] 
\centering
\includegraphics[width=5in]{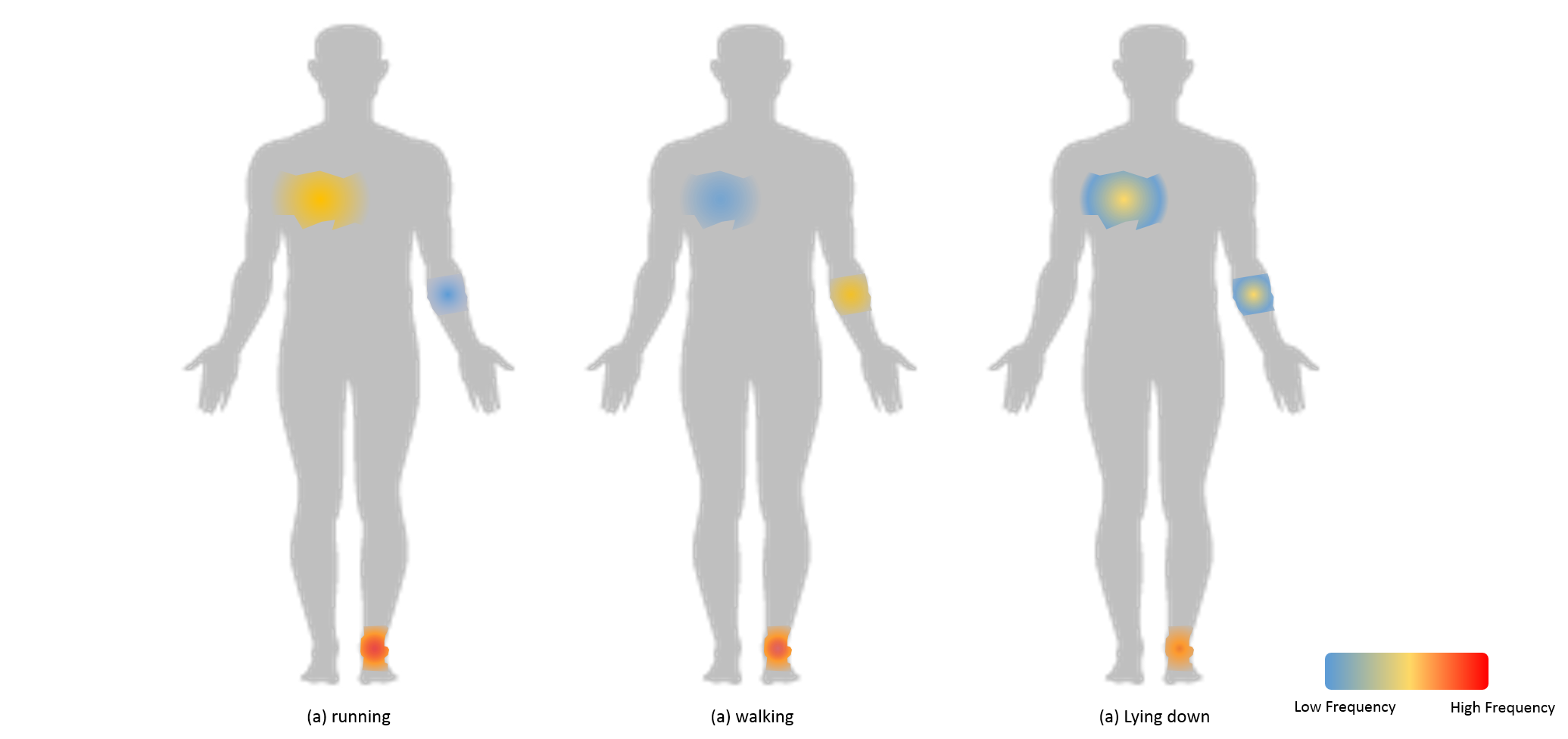} 
\caption{Glimpse Heatmap}
\label{fig: Glimpse Heatmap} 
\vspace{-0.5cm}
\end{figure}

One of the merits of our method is its interpretability. For wearable sensor-based activity recognition, subjects usually wear more than one sensors on their dominant body parts like arms, chest, and ankles, each sensor with multimodal. Attention mechanisms provide a superiority that it feeds the glimpse location back at each time step. Owing to the particularity of the activity frames, the attention model in our scenario not only provides the specific body parts it focuses on  but also highlights the most contributing sensors and modals to diverse activities. In this section, we only present the experimental results of running, walking and lying down on MHEALTH dataset for simplicity. The available sensors on MHEALTH include ECG, chest accelerometer, ankle accelerometer, ankle gyroscope, ankle magnetometer, arm accelerometer, arm gyroscope and arm magnetometer. Figure~\ref{fig: Glimpse Heatmap} shows the glimpse heatmap for all sensors. Taking running as an example, we can observe ankle as the most active part of running. The chest also contributes a lot while arm involves the least. To further demonstrate the involvement of all sensors modal data, Table~\ref{Tab: Modals Involvements on MHEALTH Dataset} concludes the percentage of our model "looking at" different modals for the latest 120 times (out of 200 times). It shows that for running, the most salient modal is ankle acceleration, which accounts for 30.55\%. ECG and ankle gyroscope data are also significant. The experimental results totally conform to the reality that while running, the most active body parts should be legs and ankles. Another self-evident truth is that in our experiments, one modal that can easily distinguish strenuous exercise like running from others is ECG. Also, since the model still "looks at" other modals for several times, it is able to better corroborate the claim that our model minimizes information loss.

\begin{table}[]
\centering
\caption{Modals Involvements on MHEALTH Dataset (\%). (Acc, Gyro, Magn denote Accelerometer, Gyroscope and Magnetometer, respectively.)}
\label{Tab: Modals Involvements on MHEALTH Dataset}
\begin{tabular}{ccccccccc}
\hline
activity&ECG & $Acc_{chest}$ & $Acc_{ankle}$ & $Gyro_{ankle}$ & $Magn_{ankle}$ & $Acc_{arm}$ & $Gyro_{arm}$ & $Magn_{arm}$\\ \hline
running & \textbf{21.98} & 10.22 & \textbf{30.55} & \textbf{15.86} & 4.30 & 6.49 & 5.03 & 5.57\\ 
walking & 7.23 & 11.05 & \textbf{18.78} & \textbf{19.26} & 8.72 & \textbf{19.66} & 9.46 & 5.83\\
lying down & 6.58 & 13.45 & \textbf{16.34} & 10.23 & \textbf{17.92} & 10.29 & 10.72 & \textbf{14.47} \\

\hline
\end{tabular}
\end{table}

\subsection{Labeled Data Dependency}

\begin{figure}[htbp] 
\centering
\includegraphics[width=6.0in]{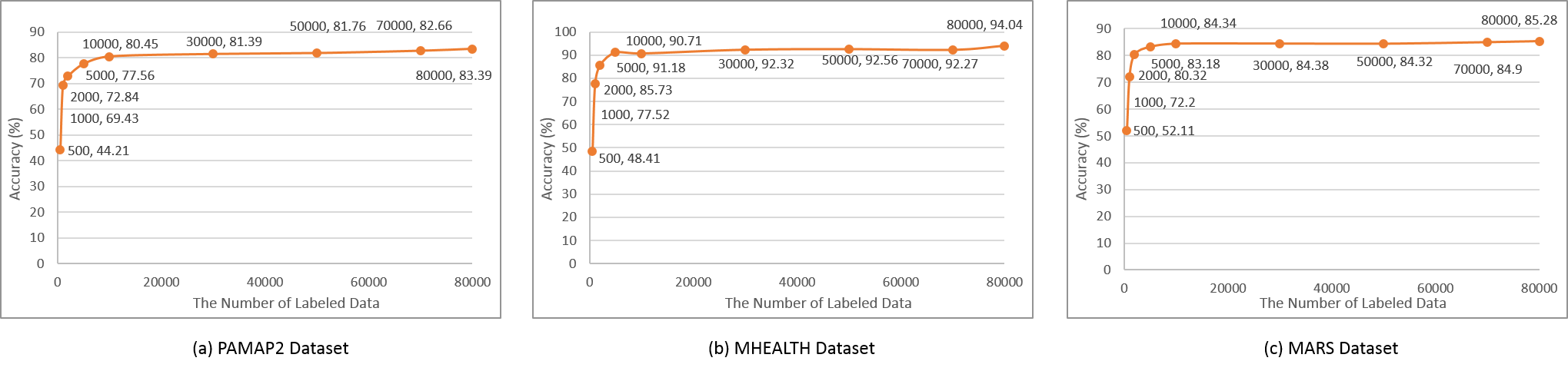} 
\caption{Labeled Data Dependency}
\label{fig: labeled data dependency} 
\vspace{-0.6cm}
\end{figure}

It is generally regarded as one of the most serious challenges in human activity recognition to get enough labeled data, owing to the considerable annotation expense and the possibility of user privacy violation. Semi-supervised \cite{yao2016learning} or weakly-supervised methods \cite{lavania2016weakly} may take advantages of unlabeled data but meanwhile incur extra cost \cite{yao2016learning}. In contrast, we propose to maximize the utilization of features and achieve the best performance with least cost. With activity frames fully extracting information among features, attention model focusing on the most salient data, and frame based recurrent network detailedly studying the temporal pattern, RAAF is able to reduce the dependency on labeled data significantly. As figure~\ref{fig: labeled data dependency} shows although the accuracy decreases with less labeled data, the downtrend is slow until the number of labeled data is reduced to 1000 on both datasets. Even 5000 labeled data deliver a relatively satisfactory accuracy. Owning to the fact that the experiments adopts Leave-One-Subject-Out (LOSO) cross-validation, which means 7 subjects' data for 6 activities on PAMAP dataset, 8 subjects on MHEALTH dataset and 6 subjects for 5 activities on MARS are used for training, only 119, 104 and 166 data  are needed for each subject and each activity on PAMAP2, MHEALTH and MARS, respectively. This fact fully validates the low dependency of our method on labeled data.

\section{Conclusion}

This paper proposes an innovative human activity recognition approach, RAAF, which includes (a) a novel form of multimodal sensor features, convolutional activity frames to fully extract relations between each pair of sensors and modality data and (b) a dual-stream convolutional attention model to combine recurrent attention and recurrent activity frames. The experiments show our method outperforms the state-of-the-art methods and has low annotation dependency which indicates that it substantially reduces the requirement for labeled data. Also, the method enjoys great interpretability in spite of the non-explanation of neural networks. Furthermore, we conduct the method on a real-world dataset collected by ourselves to validate its applicability in practical situations, Given the encouraging results, achieving higher performance by RAAF for more complex HAR scenarios is promising in our future work.

\bibliographystyle{ACM-Reference-Format}
\bibliography{Attention}


\begin{thebibliography}{50}


\ifx \showCODEN    \undefined \def \showCODEN     #1{\unskip}     \fi
\ifx \showDOI      \undefined \def \showDOI       #1{#1}\fi
\ifx \showISBNx    \undefined \def \showISBNx     #1{\unskip}     \fi
\ifx \showISBNxiii \undefined \def \showISBNxiii  #1{\unskip}     \fi
\ifx \showISSN     \undefined \def \showISSN      #1{\unskip}     \fi
\ifx \showLCCN     \undefined \def \showLCCN      #1{\unskip}     \fi
\ifx \shownote     \undefined \def \shownote      #1{#1}          \fi
\ifx \showarticletitle \undefined \def \showarticletitle #1{#1}   \fi
\ifx \showURL      \undefined \def \showURL       {\relax}        \fi
\providecommand\bibfield[2]{#2}
\providecommand\bibinfo[2]{#2}
\providecommand\natexlab[1]{#1}
\providecommand\showeprint[2][]{arXiv:#2}

\bibitem[\protect\citeauthoryear{Alexe, Heess, Teh, and Ferrari}{Alexe
  et~al\mbox{.}}{2012}]%
        {alexe2012searching}
\bibfield{author}{\bibinfo{person}{Bogdan Alexe}, \bibinfo{person}{Nicolas
  Heess}, \bibinfo{person}{Yee~W Teh}, {and} \bibinfo{person}{Vittorio
  Ferrari}.} \bibinfo{year}{2012}\natexlab{}.
\newblock \showarticletitle{Searching for objects driven by context}. In
  \bibinfo{booktitle}{{\em Advances in Neural Information Processing Systems}}.
  \bibinfo{pages}{881--889}.
\newblock


\bibitem[\protect\citeauthoryear{Althloothi, Mahoor, Zhang, and
  Voyles}{Althloothi et~al\mbox{.}}{2014}]%
        {althloothi2014human}
\bibfield{author}{\bibinfo{person}{Salah Althloothi},
  \bibinfo{person}{Mohammad~H Mahoor}, \bibinfo{person}{Xiao Zhang}, {and}
  \bibinfo{person}{Richard~M Voyles}.} \bibinfo{year}{2014}\natexlab{}.
\newblock \showarticletitle{Human activity recognition using multi-features and
  multiple kernel learning}.
\newblock \bibinfo{journal}{{\em Pattern recognition\/}} \bibinfo{volume}{47},
  \bibinfo{number}{5} (\bibinfo{year}{2014}), \bibinfo{pages}{1800--1812}.
\newblock


\bibitem[\protect\citeauthoryear{Anguita, Ghio, Oneto, Parra, and
  Reyes-Ortiz}{Anguita et~al\mbox{.}}{2013}]%
        {anguita2013public}
\bibfield{author}{\bibinfo{person}{Davide Anguita}, \bibinfo{person}{Alessandro
  Ghio}, \bibinfo{person}{Luca Oneto}, \bibinfo{person}{Xavier Parra}, {and}
  \bibinfo{person}{Jorge~Luis Reyes-Ortiz}.} \bibinfo{year}{2013}\natexlab{}.
\newblock \showarticletitle{A Public Domain Dataset for Human Activity
  Recognition using Smartphones.}. In \bibinfo{booktitle}{{\em ESANN}}.
\newblock


\bibitem[\protect\citeauthoryear{Banos, Garcia, Holgado-Terriza, Damas,
  Pomares, Rojas, Saez, and Villalonga}{Banos et~al\mbox{.}}{2014}]%
        {banos2014mhealthdroid}
\bibfield{author}{\bibinfo{person}{Oresti Banos}, \bibinfo{person}{Rafael
  Garcia}, \bibinfo{person}{Juan~A Holgado-Terriza}, \bibinfo{person}{Miguel
  Damas}, \bibinfo{person}{Hector Pomares}, \bibinfo{person}{Ignacio Rojas},
  \bibinfo{person}{Alejandro Saez}, {and} \bibinfo{person}{Claudia
  Villalonga}.} \bibinfo{year}{2014}\natexlab{}.
\newblock \showarticletitle{mHealthDroid: a novel framework for agile
  development of mobile health applications}. In \bibinfo{booktitle}{{\em
  International Workshop on Ambient Assisted Living}}. Springer,
  \bibinfo{pages}{91--98}.
\newblock


\bibitem[\protect\citeauthoryear{Banos, Villalonga, Garcia, Saez, Damas,
  Holgado-Terriza, Lee, Pomares, and Rojas}{Banos et~al\mbox{.}}{2015}]%
        {banos2015design}
\bibfield{author}{\bibinfo{person}{Oresti Banos}, \bibinfo{person}{Claudia
  Villalonga}, \bibinfo{person}{Rafael Garcia}, \bibinfo{person}{Alejandro
  Saez}, \bibinfo{person}{Miguel Damas}, \bibinfo{person}{Juan~A
  Holgado-Terriza}, \bibinfo{person}{Sungyong Lee}, \bibinfo{person}{Hector
  Pomares}, {and} \bibinfo{person}{Ignacio Rojas}.}
  \bibinfo{year}{2015}\natexlab{}.
\newblock \showarticletitle{Design, implementation and validation of a novel
  open framework for agile development of mobile health applications}.
\newblock \bibinfo{journal}{{\em Biomedical engineering online\/}}
  \bibinfo{volume}{14}, \bibinfo{number}{2} (\bibinfo{year}{2015}),
  \bibinfo{pages}{S6}.
\newblock


\bibitem[\protect\citeauthoryear{Bengio}{Bengio}{2013}]%
        {bengio2013deep}
\bibfield{author}{\bibinfo{person}{Yoshua Bengio}.}
  \bibinfo{year}{2013}\natexlab{}.
\newblock \showarticletitle{Deep learning of representations: Looking forward}.
  In \bibinfo{booktitle}{{\em International Conference on Statistical Language
  and Speech Processing}}. Springer, \bibinfo{pages}{1--37}.
\newblock


\bibitem[\protect\citeauthoryear{Bulling, Blanke, and Schiele}{Bulling
  et~al\mbox{.}}{2014}]%
        {bulling2014tutorial}
\bibfield{author}{\bibinfo{person}{Andreas Bulling}, \bibinfo{person}{Ulf
  Blanke}, {and} \bibinfo{person}{Bernt Schiele}.}
  \bibinfo{year}{2014}\natexlab{}.
\newblock \showarticletitle{A tutorial on human activity recognition using
  body-worn inertial sensors}.
\newblock \bibinfo{journal}{{\em ACM Computing Surveys (CSUR)\/}}
  \bibinfo{volume}{46}, \bibinfo{number}{3} (\bibinfo{year}{2014}),
  \bibinfo{pages}{33}.
\newblock


\bibitem[\protect\citeauthoryear{Butko and Movellan}{Butko and
  Movellan}{2008}]%
        {butko2008pomdp}
\bibfield{author}{\bibinfo{person}{Nicholas~J Butko} {and}
  \bibinfo{person}{Javier~R Movellan}.} \bibinfo{year}{2008}\natexlab{}.
\newblock \showarticletitle{I-POMDP: An infomax model of eye movement}. In
  \bibinfo{booktitle}{{\em Development and Learning, 2008. ICDL 2008. 7th IEEE
  International Conference on}}. IEEE, \bibinfo{pages}{139--144}.
\newblock


\bibitem[\protect\citeauthoryear{Chen, Hoey, Nugent, Cook, and Yu}{Chen
  et~al\mbox{.}}{2012}]%
        {chen2012sensor}
\bibfield{author}{\bibinfo{person}{Liming Chen}, \bibinfo{person}{Jesse Hoey},
  \bibinfo{person}{Chris~D Nugent}, \bibinfo{person}{Diane~J Cook}, {and}
  \bibinfo{person}{Zhiwen Yu}.} \bibinfo{year}{2012}\natexlab{}.
\newblock \showarticletitle{Sensor-based activity recognition}.
\newblock \bibinfo{journal}{{\em IEEE Transactions on Systems, Man, and
  Cybernetics, Part C (Applications and Reviews)\/}} \bibinfo{volume}{42},
  \bibinfo{number}{6} (\bibinfo{year}{2012}), \bibinfo{pages}{790--808}.
\newblock


\bibitem[\protect\citeauthoryear{Deng}{Deng}{2014}]%
        {deng2014tutorial}
\bibfield{author}{\bibinfo{person}{Li Deng}.} \bibinfo{year}{2014}\natexlab{}.
\newblock \showarticletitle{A tutorial survey of architectures, algorithms, and
  applications for deep learning}.
\newblock \bibinfo{journal}{{\em APSIPA Transactions on Signal and Information
  Processing\/}}  \bibinfo{volume}{3} (\bibinfo{year}{2014}).
\newblock


\bibitem[\protect\citeauthoryear{Denil, Bazzani, Larochelle, and
  de~Freitas}{Denil et~al\mbox{.}}{2012}]%
        {denil2012learning}
\bibfield{author}{\bibinfo{person}{Misha Denil}, \bibinfo{person}{Loris
  Bazzani}, \bibinfo{person}{Hugo Larochelle}, {and} \bibinfo{person}{Nando de
  Freitas}.} \bibinfo{year}{2012}\natexlab{}.
\newblock \showarticletitle{Learning where to attend with deep architectures
  for image tracking}.
\newblock \bibinfo{journal}{{\em Neural computation\/}} \bibinfo{volume}{24},
  \bibinfo{number}{8} (\bibinfo{year}{2012}), \bibinfo{pages}{2151--2184}.
\newblock


\bibitem[\protect\citeauthoryear{Edel and K{\"o}ppe}{Edel and
  K{\"o}ppe}{2016}]%
        {edel2016binarized}
\bibfield{author}{\bibinfo{person}{Marcus Edel} {and} \bibinfo{person}{Enrico
  K{\"o}ppe}.} \bibinfo{year}{2016}\natexlab{}.
\newblock \showarticletitle{Binarized-BLSTM-RNN based Human Activity
  Recognition}. In \bibinfo{booktitle}{{\em Indoor Positioning and Indoor
  Navigation (IPIN), 2016 International Conference on}}. IEEE,
  \bibinfo{pages}{1--7}.
\newblock


\bibitem[\protect\citeauthoryear{Fang and Hu}{Fang and Hu}{2014}]%
        {fang2014recognizing}
\bibfield{author}{\bibinfo{person}{Hongqing Fang} {and} \bibinfo{person}{Chen
  Hu}.} \bibinfo{year}{2014}\natexlab{}.
\newblock \showarticletitle{Recognizing human activity in smart home using deep
  learning algorithm}. In \bibinfo{booktitle}{{\em Control Conference (CCC),
  2014 33rd Chinese}}. IEEE, \bibinfo{pages}{4716--4720}.
\newblock


\bibitem[\protect\citeauthoryear{Guan and Pl\"{o}tz}{Guan and
  Pl\"{o}tz}{2017}]%
        {Guan:2017:EDL:3120957.3090076}
\bibfield{author}{\bibinfo{person}{Yu Guan} {and} \bibinfo{person}{Thomas
  Pl\"{o}tz}.} \bibinfo{year}{2017}\natexlab{}.
\newblock \showarticletitle{Ensembles of Deep LSTM Learners for Activity
  Recognition Using Wearables}.
\newblock \bibinfo{journal}{{\em Proc. ACM Interact. Mob. Wearable Ubiquitous
  Technol.\/}} \bibinfo{volume}{1}, \bibinfo{number}{2}, Article
  \bibinfo{articleno}{11} (\bibinfo{date}{June} \bibinfo{year}{2017}),
  \bibinfo{numpages}{28}~pages.
\newblock
\showISSN{2474-9567}
\showDOI{%
\url{https://doi.org/10.1145/3090076}}


\bibitem[\protect\citeauthoryear{Guo, Chen, Peng, and Chen}{Guo
  et~al\mbox{.}}{2016}]%
        {guo2016wearable}
\bibfield{author}{\bibinfo{person}{Haodong Guo}, \bibinfo{person}{Ling Chen},
  \bibinfo{person}{Liangying Peng}, {and} \bibinfo{person}{Gencai Chen}.}
  \bibinfo{year}{2016}\natexlab{}.
\newblock \showarticletitle{Wearable sensor based multimodal human activity
  recognition exploiting the diversity of classifier ensemble}. In
  \bibinfo{booktitle}{{\em Proceedings of the 2016 ACM International Joint
  Conference on Pervasive and Ubiquitous Computing}}. ACM,
  \bibinfo{pages}{1112--1123}.
\newblock


\bibitem[\protect\citeauthoryear{Guo, Chen, Shen, and Chen}{Guo
  et~al\mbox{.}}{2014}]%
        {guo2014activity}
\bibfield{author}{\bibinfo{person}{Haodong Guo}, \bibinfo{person}{Ling Chen},
  \bibinfo{person}{Yanbin Shen}, {and} \bibinfo{person}{Gencai Chen}.}
  \bibinfo{year}{2014}\natexlab{}.
\newblock \showarticletitle{Activity recognition exploiting classifier level
  fusion of acceleration and physiological signals}. In
  \bibinfo{booktitle}{{\em Proceedings of the 2014 ACM International Joint
  Conference on Pervasive and Ubiquitous Computing: Adjunct Publication}}. ACM,
  \bibinfo{pages}{63--66}.
\newblock


\bibitem[\protect\citeauthoryear{Ha, Yun, and Choi}{Ha et~al\mbox{.}}{2015}]%
        {ha2015multi}
\bibfield{author}{\bibinfo{person}{Sojeong Ha}, \bibinfo{person}{Jeong-Min
  Yun}, {and} \bibinfo{person}{Seungjin Choi}.}
  \bibinfo{year}{2015}\natexlab{}.
\newblock \showarticletitle{Multi-modal convolutional neural networks for
  activity recognition}. In \bibinfo{booktitle}{{\em Systems, Man, and
  Cybernetics (SMC), 2015 IEEE International Conference on}}. IEEE,
  \bibinfo{pages}{3017--3022}.
\newblock


\bibitem[\protect\citeauthoryear{Haque, Alahi, and Fei-Fei}{Haque
  et~al\mbox{.}}{2016}]%
        {haque2016recurrent}
\bibfield{author}{\bibinfo{person}{Albert Haque}, \bibinfo{person}{Alexandre
  Alahi}, {and} \bibinfo{person}{Li Fei-Fei}.} \bibinfo{year}{2016}\natexlab{}.
\newblock \showarticletitle{Recurrent attention models for depth-based person
  identification}. In \bibinfo{booktitle}{{\em Proceedings of the IEEE
  Conference on Computer Vision and Pattern Recognition}}.
  \bibinfo{pages}{1229--1238}.
\newblock


\bibitem[\protect\citeauthoryear{Hinton, Osindero, and Teh}{Hinton
  et~al\mbox{.}}{2006}]%
        {hinton2006fast}
\bibfield{author}{\bibinfo{person}{Geoffrey~E Hinton}, \bibinfo{person}{Simon
  Osindero}, {and} \bibinfo{person}{Yee-Whye Teh}.}
  \bibinfo{year}{2006}\natexlab{}.
\newblock \showarticletitle{A fast learning algorithm for deep belief nets}.
\newblock \bibinfo{journal}{{\em Neural computation\/}} \bibinfo{volume}{18},
  \bibinfo{number}{7} (\bibinfo{year}{2006}), \bibinfo{pages}{1527--1554}.
\newblock


\bibitem[\protect\citeauthoryear{Inoue, Inoue, and Nishida}{Inoue
  et~al\mbox{.}}{2016}]%
        {inoue2016deep}
\bibfield{author}{\bibinfo{person}{Masaya Inoue}, \bibinfo{person}{Sozo Inoue},
  {and} \bibinfo{person}{Takeshi Nishida}.} \bibinfo{year}{2016}\natexlab{}.
\newblock \showarticletitle{Deep Recurrent Neural Network for Mobile Human
  Activity Recognition with High Throughput}.
\newblock \bibinfo{journal}{{\em arXiv preprint arXiv:1611.03607\/}}
  (\bibinfo{year}{2016}).
\newblock


\bibitem[\protect\citeauthoryear{Jiang and Yin}{Jiang and Yin}{2015}]%
        {jiang2015human}
\bibfield{author}{\bibinfo{person}{Wenchao Jiang} {and}
  \bibinfo{person}{Zhaozheng Yin}.} \bibinfo{year}{2015}\natexlab{}.
\newblock \showarticletitle{Human activity recognition using wearable sensors
  by deep convolutional neural networks}. In \bibinfo{booktitle}{{\em
  Proceedings of the 23rd ACM international conference on Multimedia}}. ACM,
  \bibinfo{pages}{1307--1310}.
\newblock


\bibitem[\protect\citeauthoryear{Kavukcuoglu, Sermanet, Boureau, Gregor,
  Mathieu, and Cun}{Kavukcuoglu et~al\mbox{.}}{2010}]%
        {kavukcuoglu2010learning}
\bibfield{author}{\bibinfo{person}{Koray Kavukcuoglu}, \bibinfo{person}{Pierre
  Sermanet}, \bibinfo{person}{Y-Lan Boureau}, \bibinfo{person}{Karol Gregor},
  \bibinfo{person}{Micha{\"e}l Mathieu}, {and} \bibinfo{person}{Yann~L Cun}.}
  \bibinfo{year}{2010}\natexlab{}.
\newblock \showarticletitle{Learning convolutional feature hierarchies for
  visual recognition}. In \bibinfo{booktitle}{{\em Advances in neural
  information processing systems}}. \bibinfo{pages}{1090--1098}.
\newblock


\bibitem[\protect\citeauthoryear{Kunze and Lukowicz}{Kunze and
  Lukowicz}{2008}]%
        {kunze2008dealing}
\bibfield{author}{\bibinfo{person}{Kai Kunze} {and} \bibinfo{person}{Paul
  Lukowicz}.} \bibinfo{year}{2008}\natexlab{}.
\newblock \showarticletitle{Dealing with sensor displacement in motion-based
  onbody activity recognition systems}. In \bibinfo{booktitle}{{\em Proceedings
  of the 10th international conference on Ubiquitous computing}}. ACM,
  \bibinfo{pages}{20--29}.
\newblock


\bibitem[\protect\citeauthoryear{Lai, Xu, Chen, Yang, and Zhang}{Lai
  et~al\mbox{.}}{2014}]%
        {lai2014multilinear}
\bibfield{author}{\bibinfo{person}{Zhihui Lai}, \bibinfo{person}{Yong Xu},
  \bibinfo{person}{Qingcai Chen}, \bibinfo{person}{Jian Yang}, {and}
  \bibinfo{person}{David Zhang}.} \bibinfo{year}{2014}\natexlab{}.
\newblock \showarticletitle{Multilinear sparse principal component analysis}.
\newblock \bibinfo{journal}{{\em IEEE transactions on neural networks and
  learning systems\/}} \bibinfo{volume}{25}, \bibinfo{number}{10}
  (\bibinfo{year}{2014}), \bibinfo{pages}{1942--1950}.
\newblock


\bibitem[\protect\citeauthoryear{Lane, Georgiev, and Qendro}{Lane
  et~al\mbox{.}}{2015}]%
        {lane2015deepear}
\bibfield{author}{\bibinfo{person}{Nicholas~D Lane}, \bibinfo{person}{Petko
  Georgiev}, {and} \bibinfo{person}{Lorena Qendro}.}
  \bibinfo{year}{2015}\natexlab{}.
\newblock \showarticletitle{Deepear: robust smartphone audio sensing in
  unconstrained acoustic environments using deep learning}. In
  \bibinfo{booktitle}{{\em Proceedings of the 2015 ACM International Joint
  Conference on Pervasive and Ubiquitous Computing}}. ACM,
  \bibinfo{pages}{283--294}.
\newblock


\bibitem[\protect\citeauthoryear{Lara, P{\'e}rez, Labrador, and Posada}{Lara
  et~al\mbox{.}}{2012}]%
        {lara2012centinela}
\bibfield{author}{\bibinfo{person}{Oscar~D Lara}, \bibinfo{person}{Alfredo~J
  P{\'e}rez}, \bibinfo{person}{Miguel~A Labrador}, {and}
  \bibinfo{person}{Jos{\'e}~D Posada}.} \bibinfo{year}{2012}\natexlab{}.
\newblock \showarticletitle{Centinela: A human activity recognition system
  based on acceleration and vital sign data}.
\newblock \bibinfo{journal}{{\em Pervasive and mobile computing\/}}
  \bibinfo{volume}{8}, \bibinfo{number}{5} (\bibinfo{year}{2012}),
  \bibinfo{pages}{717--729}.
\newblock


\bibitem[\protect\citeauthoryear{Larochelle and Hinton}{Larochelle and
  Hinton}{2010}]%
        {larochelle2010learning}
\bibfield{author}{\bibinfo{person}{Hugo Larochelle} {and}
  \bibinfo{person}{Geoffrey~E Hinton}.} \bibinfo{year}{2010}\natexlab{}.
\newblock \showarticletitle{Learning to combine foveal glimpses with a
  third-order Boltzmann machine}. In \bibinfo{booktitle}{{\em Advances in
  neural information processing systems}}. \bibinfo{pages}{1243--1251}.
\newblock


\bibitem[\protect\citeauthoryear{Lavania, Thulasidasan, LaMarca, Scofield, and
  Bilmes}{Lavania et~al\mbox{.}}{2016}]%
        {lavania2016weakly}
\bibfield{author}{\bibinfo{person}{Chandrashekhar Lavania},
  \bibinfo{person}{Sunil Thulasidasan}, \bibinfo{person}{Anthony LaMarca},
  \bibinfo{person}{Jeffrey Scofield}, {and} \bibinfo{person}{Jeff Bilmes}.}
  \bibinfo{year}{2016}\natexlab{}.
\newblock \showarticletitle{A weakly supervised activity recognition framework
  for real-time synthetic biology laboratory assistance}. In
  \bibinfo{booktitle}{{\em Proceedings of the 2016 ACM International Joint
  Conference on Pervasive and Ubiquitous Computing}}. ACM,
  \bibinfo{pages}{37--48}.
\newblock


\bibitem[\protect\citeauthoryear{LeCun, Bengio, and Hinton}{LeCun
  et~al\mbox{.}}{2015}]%
        {lecun2015deep}
\bibfield{author}{\bibinfo{person}{Yann LeCun}, \bibinfo{person}{Yoshua
  Bengio}, {and} \bibinfo{person}{Geoffrey Hinton}.}
  \bibinfo{year}{2015}\natexlab{}.
\newblock \showarticletitle{Deep learning}.
\newblock \bibinfo{journal}{{\em Nature\/}} \bibinfo{volume}{521},
  \bibinfo{number}{7553} (\bibinfo{year}{2015}), \bibinfo{pages}{436--444}.
\newblock


\bibitem[\protect\citeauthoryear{Li, Shi, Ding, and Liu}{Li
  et~al\mbox{.}}{2014}]%
        {li2014unsupervised}
\bibfield{author}{\bibinfo{person}{Yongmou Li}, \bibinfo{person}{Dianxi Shi},
  \bibinfo{person}{Bo Ding}, {and} \bibinfo{person}{Dongbo Liu}.}
  \bibinfo{year}{2014}\natexlab{}.
\newblock \showarticletitle{Unsupervised feature learning for human activity
  recognition using smartphone sensors}.
\newblock In \bibinfo{booktitle}{{\em Mining Intelligence and Knowledge
  Exploration}}. \bibinfo{publisher}{Springer}, \bibinfo{pages}{99--107}.
\newblock


\bibitem[\protect\citeauthoryear{Mnih, Heess, Graves, et~al\mbox{.}}{Mnih
  et~al\mbox{.}}{2014}]%
        {mnih2014recurrent}
\bibfield{author}{\bibinfo{person}{Volodymyr Mnih}, \bibinfo{person}{Nicolas
  Heess}, \bibinfo{person}{Alex Graves}, {et~al\mbox{.}}}
  \bibinfo{year}{2014}\natexlab{}.
\newblock \showarticletitle{Recurrent models of visual attention}. In
  \bibinfo{booktitle}{{\em Advances in neural information processing systems}}.
  \bibinfo{pages}{2204--2212}.
\newblock


\bibitem[\protect\citeauthoryear{Ooi, Teoh, Pang, and Hiew}{Ooi
  et~al\mbox{.}}{2016}]%
        {ooi2016image}
\bibfield{author}{\bibinfo{person}{Shih~Yin Ooi}, \bibinfo{person}{Andrew
  Beng~Jin Teoh}, \bibinfo{person}{Ying~Han Pang}, {and}
  \bibinfo{person}{Bee~Yan Hiew}.} \bibinfo{year}{2016}\natexlab{}.
\newblock \showarticletitle{Image-based handwritten signature verification
  using hybrid methods of discrete radon transform, principal component
  analysis and probabilistic neural network}.
\newblock \bibinfo{journal}{{\em Applied Soft Computing\/}}
  \bibinfo{volume}{40} (\bibinfo{year}{2016}), \bibinfo{pages}{274--282}.
\newblock


\bibitem[\protect\citeauthoryear{Parkka, Ermes, Korpipaa, Mantyjarvi, Peltola,
  and Korhonen}{Parkka et~al\mbox{.}}{2006}]%
        {parkka2006activity}
\bibfield{author}{\bibinfo{person}{Juha Parkka}, \bibinfo{person}{Miikka
  Ermes}, \bibinfo{person}{Panu Korpipaa}, \bibinfo{person}{Jani Mantyjarvi},
  \bibinfo{person}{Johannes Peltola}, {and} \bibinfo{person}{Ilkka Korhonen}.}
  \bibinfo{year}{2006}\natexlab{}.
\newblock \showarticletitle{Activity classification using realistic data from
  wearable sensors}.
\newblock \bibinfo{journal}{{\em IEEE Transactions on information technology in
  biomedicine\/}} \bibinfo{volume}{10}, \bibinfo{number}{1}
  (\bibinfo{year}{2006}), \bibinfo{pages}{119--128}.
\newblock


\bibitem[\protect\citeauthoryear{Phidgets}{Phidgets}{2010}]%
        {phidgets20101056}
\bibfield{author}{\bibinfo{person}{I Phidgets}.}
  \bibinfo{year}{2010}\natexlab{}.
\newblock \showarticletitle{1056-PhidgetSpatial 3/3/3}.
\newblock \bibinfo{journal}{{\em Code Samples For This Product\/}}
  (\bibinfo{year}{2010}).
\newblock


\bibitem[\protect\citeauthoryear{Pl{\"o}tz, Hammerla, and Olivier}{Pl{\"o}tz
  et~al\mbox{.}}{2011}]%
        {plotz2011feature}
\bibfield{author}{\bibinfo{person}{Thomas Pl{\"o}tz}, \bibinfo{person}{Nils~Y
  Hammerla}, {and} \bibinfo{person}{Patrick Olivier}.}
  \bibinfo{year}{2011}\natexlab{}.
\newblock \showarticletitle{Feature learning for activity recognition in
  ubiquitous computing}. In \bibinfo{booktitle}{{\em IJCAI
  Proceedings-International Joint Conference on Artificial Intelligence}},
  Vol.~\bibinfo{volume}{22}. \bibinfo{pages}{1729}.
\newblock


\bibitem[\protect\citeauthoryear{Radu, Lane, Bhattacharya, Mascolo, Marina, and
  Kawsar}{Radu et~al\mbox{.}}{2016}]%
        {radu2016towards}
\bibfield{author}{\bibinfo{person}{Valentin Radu}, \bibinfo{person}{Nicholas~D
  Lane}, \bibinfo{person}{Sourav Bhattacharya}, \bibinfo{person}{Cecilia
  Mascolo}, \bibinfo{person}{Mahesh~K Marina}, {and} \bibinfo{person}{Fahim
  Kawsar}.} \bibinfo{year}{2016}\natexlab{}.
\newblock \showarticletitle{Towards multimodal deep learning for activity
  recognition on mobile devices}. In \bibinfo{booktitle}{{\em Proceedings of
  the 2016 ACM International Joint Conference on Pervasive and Ubiquitous
  Computing: Adjunct}}. ACM, \bibinfo{pages}{185--188}.
\newblock


\bibitem[\protect\citeauthoryear{Reiss and Stricker}{Reiss and
  Stricker}{2012a}]%
        {reiss2012creating}
\bibfield{author}{\bibinfo{person}{Attila Reiss} {and} \bibinfo{person}{Didier
  Stricker}.} \bibinfo{year}{2012}\natexlab{a}.
\newblock \showarticletitle{Creating and benchmarking a new dataset for
  physical activity monitoring}. In \bibinfo{booktitle}{{\em Proceedings of the
  5th International Conference on PErvasive Technologies Related to Assistive
  Environments}}. ACM, \bibinfo{pages}{40}.
\newblock


\bibitem[\protect\citeauthoryear{Reiss and Stricker}{Reiss and
  Stricker}{2012b}]%
        {reiss2012introducing}
\bibfield{author}{\bibinfo{person}{Attila Reiss} {and} \bibinfo{person}{Didier
  Stricker}.} \bibinfo{year}{2012}\natexlab{b}.
\newblock \showarticletitle{Introducing a new benchmarked dataset for activity
  monitoring}. In \bibinfo{booktitle}{{\em Wearable Computers (ISWC), 2012 16th
  International Symposium on}}. IEEE, \bibinfo{pages}{108--109}.
\newblock


\bibitem[\protect\citeauthoryear{Singh, Pondenkandath, Zhou, Lukowicz, and
  Liwicki}{Singh et~al\mbox{.}}{2017}]%
        {singh2017transforming}
\bibfield{author}{\bibinfo{person}{Monit~Shah Singh},
  \bibinfo{person}{Vinaychandran Pondenkandath}, \bibinfo{person}{Bo Zhou},
  \bibinfo{person}{Paul Lukowicz}, {and} \bibinfo{person}{Marcus Liwicki}.}
  \bibinfo{year}{2017}\natexlab{}.
\newblock \showarticletitle{Transforming Sensor Data to the Image Domain for
  Deep Learning-an Application to Footstep Detection}.
\newblock \bibinfo{journal}{{\em Neural Networks (IJCNN), 2017 International
  Joint Conference on\/}} (\bibinfo{year}{2017}), \bibinfo{pages}{3017--3022}.
\newblock


\bibitem[\protect\citeauthoryear{Tapia, Intille, Haskell, Larson, Wright, King,
  and Friedman}{Tapia et~al\mbox{.}}{2007}]%
        {tapia2007real}
\bibfield{author}{\bibinfo{person}{Emmanuel~Munguia Tapia},
  \bibinfo{person}{Stephen~S Intille}, \bibinfo{person}{William Haskell},
  \bibinfo{person}{Kent Larson}, \bibinfo{person}{Julie Wright},
  \bibinfo{person}{Abby King}, {and} \bibinfo{person}{Robert Friedman}.}
  \bibinfo{year}{2007}\natexlab{}.
\newblock \showarticletitle{Real-time recognition of physical activities and
  their intensities using wireless accelerometers and a heart rate monitor}. In
  \bibinfo{booktitle}{{\em Wearable Computers, 2007 11th IEEE International
  Symposium on}}. IEEE, \bibinfo{pages}{37--40}.
\newblock


\bibitem[\protect\citeauthoryear{Wang, Chen, Shang, Zhang, and Liu}{Wang
  et~al\mbox{.}}{2016}]%
        {wang2016human}
\bibfield{author}{\bibinfo{person}{Aiguo Wang}, \bibinfo{person}{Guilin Chen},
  \bibinfo{person}{Cuijuan Shang}, \bibinfo{person}{Miaofei Zhang}, {and}
  \bibinfo{person}{Li Liu}.} \bibinfo{year}{2016}\natexlab{}.
\newblock \showarticletitle{Human Activity Recognition in a Smart Home
  Environment with Stacked Denoising Autoencoders}. In \bibinfo{booktitle}{{\em
  International Conference on Web-Age Information Management}}. Springer,
  \bibinfo{pages}{29--40}.
\newblock


\bibitem[\protect\citeauthoryear{Wang and Wang}{Wang and Wang}{2017}]%
        {wang2017modeling}
\bibfield{author}{\bibinfo{person}{Hongsong Wang} {and} \bibinfo{person}{Liang
  Wang}.} \bibinfo{year}{2017}\natexlab{}.
\newblock \showarticletitle{Modeling Temporal Dynamics and Spatial
  Configurations of Actions Using Two-Stream Recurrent Neural Networks}.
\newblock \bibinfo{journal}{{\em The Conference on Computer Vision and Pattern
  Recognition (CVPR)\/}} (\bibinfo{year}{2017}).
\newblock


\bibitem[\protect\citeauthoryear{Williams}{Williams}{1992}]%
        {williams1992simple}
\bibfield{author}{\bibinfo{person}{Ronald~J Williams}.}
  \bibinfo{year}{1992}\natexlab{}.
\newblock \showarticletitle{Simple statistical gradient-following algorithms
  for connectionist reinforcement learning}.
\newblock \bibinfo{journal}{{\em Machine learning\/}} \bibinfo{volume}{8},
  \bibinfo{number}{3-4} (\bibinfo{year}{1992}), \bibinfo{pages}{229--256}.
\newblock


\bibitem[\protect\citeauthoryear{Wold, Esbensen, and Geladi}{Wold
  et~al\mbox{.}}{1987}]%
        {wold1987principal}
\bibfield{author}{\bibinfo{person}{Svante Wold}, \bibinfo{person}{Kim
  Esbensen}, {and} \bibinfo{person}{Paul Geladi}.}
  \bibinfo{year}{1987}\natexlab{}.
\newblock \showarticletitle{Principal component analysis}.
\newblock \bibinfo{journal}{{\em Chemometrics and intelligent laboratory
  systems\/}} \bibinfo{volume}{2}, \bibinfo{number}{1-3}
  (\bibinfo{year}{1987}), \bibinfo{pages}{37--52}.
\newblock


\bibitem[\protect\citeauthoryear{Yacoob and Black}{Yacoob and Black}{1998}]%
        {yacoob1998parameterized}
\bibfield{author}{\bibinfo{person}{Yaser Yacoob} {and}
  \bibinfo{person}{Michael~J Black}.} \bibinfo{year}{1998}\natexlab{}.
\newblock \showarticletitle{Parameterized modeling and recognition of
  activities}. In \bibinfo{booktitle}{{\em Computer Vision, 1998. Sixth
  International Conference on}}. IEEE, \bibinfo{pages}{120--127}.
\newblock


\bibitem[\protect\citeauthoryear{Yang, Nguyen, San, Li, and Krishnaswamy}{Yang
  et~al\mbox{.}}{2015}]%
        {yang2015deep}
\bibfield{author}{\bibinfo{person}{Jianbo Yang}, \bibinfo{person}{Minh~Nhut
  Nguyen}, \bibinfo{person}{Phyo~Phyo San}, \bibinfo{person}{Xiaoli Li}, {and}
  \bibinfo{person}{Shonali Krishnaswamy}.} \bibinfo{year}{2015}\natexlab{}.
\newblock \showarticletitle{Deep Convolutional Neural Networks on Multichannel
  Time Series for Human Activity Recognition.}. In \bibinfo{booktitle}{{\em
  IJCAI}}. \bibinfo{pages}{3995--4001}.
\newblock


\bibitem[\protect\citeauthoryear{Yao, Nie, Sheng, Gu, Li, and Wang}{Yao
  et~al\mbox{.}}{2016}]%
        {yao2016learning}
\bibfield{author}{\bibinfo{person}{Lina Yao}, \bibinfo{person}{Feiping Nie},
  \bibinfo{person}{Quan~Z Sheng}, \bibinfo{person}{Tao Gu},
  \bibinfo{person}{Xue Li}, {and} \bibinfo{person}{Sen Wang}.}
  \bibinfo{year}{2016}\natexlab{}.
\newblock \showarticletitle{Learning from less for better: semi-supervised
  activity recognition via shared structure discovery}. In
  \bibinfo{booktitle}{{\em Proceedings of the 2016 ACM International Joint
  Conference on Pervasive and Ubiquitous Computing}}. ACM,
  \bibinfo{pages}{13--24}.
\newblock


\bibitem[\protect\citeauthoryear{Zeng, Nguyen, Yu, Mengshoel, Zhu, Wu, and
  Zhang}{Zeng et~al\mbox{.}}{2014}]%
        {zeng2014convolutional}
\bibfield{author}{\bibinfo{person}{Ming Zeng}, \bibinfo{person}{Le~T Nguyen},
  \bibinfo{person}{Bo Yu}, \bibinfo{person}{Ole~J Mengshoel},
  \bibinfo{person}{Jiang Zhu}, \bibinfo{person}{Pang Wu}, {and}
  \bibinfo{person}{Joy Zhang}.} \bibinfo{year}{2014}\natexlab{}.
\newblock \showarticletitle{Convolutional neural networks for human activity
  recognition using mobile sensors}. In \bibinfo{booktitle}{{\em Mobile
  Computing, Applications and Services (MobiCASE), 2014 6th International
  Conference on}}. IEEE, \bibinfo{pages}{197--205}.
\newblock


\bibitem[\protect\citeauthoryear{Zhang and Sawchuk}{Zhang and Sawchuk}{2012}]%
        {zhang2012usc}
\bibfield{author}{\bibinfo{person}{Mi Zhang} {and} \bibinfo{person}{Alexander~A
  Sawchuk}.} \bibinfo{year}{2012}\natexlab{}.
\newblock \showarticletitle{USC-HAD: a daily activity dataset for ubiquitous
  activity recognition using wearable sensors}. In \bibinfo{booktitle}{{\em
  Proceedings of the 2012 ACM Conference on Ubiquitous Computing}}. ACM,
  \bibinfo{pages}{1036--1043}.
\newblock


\bibitem[\protect\citeauthoryear{Zontak, Mosseri, and Irani}{Zontak
  et~al\mbox{.}}{2013}]%
        {zontak2013separating}
\bibfield{author}{\bibinfo{person}{Maria Zontak}, \bibinfo{person}{Inbar
  Mosseri}, {and} \bibinfo{person}{Michal Irani}.}
  \bibinfo{year}{2013}\natexlab{}.
\newblock \showarticletitle{Separating signal from noise using patch recurrence
  across scales}. In \bibinfo{booktitle}{{\em Proceedings of the IEEE
  Conference on Computer Vision and Pattern Recognition}}.
  \bibinfo{pages}{1195--1202}.
\newblock


\end{thebibliography}

\end{document}